\documentclass[aps,prb,twocolumn,superscriptaddress,showpacs]{revtex4-1}
\pdfoutput=1

\usepackage{epstopdf}
\usepackage{epsfig}
\usepackage{amsmath}
\usepackage{amssymb}
\usepackage[colorlinks=true,citecolor=blue,linkcolor=blue]{hyperref}

\begin{document}

\title{Interaction-driven topological insulators on the kagome and the decorated honeycomb lattices}

\author{Jun Wen}
\author{Andreas R\"uegg}
\affiliation{Department of Physics, The University of Texas at Austin, Austin, Texas 78712, USA}
\author{C.-C. Joseph Wang}
\affiliation{Department of Physics, Georgetown University, 37th and O Sts. NW, Washington, DC 20057, USA}
\author{Gregory A. Fiete}
\affiliation{Department of Physics, The University of Texas at Austin, Austin, Texas 78712, USA}

\date{\today}

\begin{abstract}
We study the spinless and spinful extended Hubbard models with repulsive interactions on the kagome and the decorated honeycomb (``star") lattice. Using Hartree-Fock mean-field theory, we show that interaction-driven insulating phases with non-trivial topological invariants (Chern number or $Z_2$ invariant) exist for an experimentally reasonable range of parameters. These phases occur at filling fractions which involve either Dirac points or quadratic band crossing points in the non-interacting limit. We present comprehensive mean-field phase diagrams for these lattices and discuss the competition between topologically non-trivial phases and numerous other ordered states, including various charge, spin, and bond orderings.  Our results suggest that $Z_2$ topological insulators should be found in a number of systems with either little or no intrinsic spin-orbit coupling.
\end{abstract}



\maketitle
\section{Introduction}

The study of topological properties of a quantum system with many
degrees of freedom can provide insights into global features of
ground states and can reveal physical behaviors which are robust
against weak perturbations. While the terminology of {\it topological order} has been used to describe different aspects of a quantum system with interactions,\cite{WenBook:2004,Nayak:rmp08,Volovik} we will focus on topological properties which are stored in the set of single-particle wave functions describing band structures of materials with a bulk gap. One famous example is the integer quantum Hall effect where the topological property is encapsulated in an integer called the {\it Chern number}. It has been shown that the Chern number is directly related to quantized values of the Hall conductivity.\cite{Thouless:1982} A nonzero Chern number requires the breaking of time-reversal symmetry either by an external magnetic field or, in the absent of a net magnetic flux through a unit cell, by microscopically circling currents.\cite{Haldane:1988}

Breaking of time-reversal symmetry is not necessarily required to define topological invariants which distinguish different bulk insulators (or superconductors). Based on the random matrix theory, a comprehensive classification scheme for non-interacting systems has been worked out.\cite{Schnyder:2008} Among all classifications, {\it topological insulators} (TIs) with time-reversal symmetry have raised considerable interest in recent years (see Refs.~[\onlinecite{Moore:2010,Kane:2010,Zhang:2010}]). TIs are well described by conventional band theory.  However, they are a distinct phase of matter with bulk energy gaps and an odd number of time-reversal symmetry protected gapless modes on their edge (surface in three dimensions).\cite{Kane:2005a,Kane:2005b,Kane:2007} In two dimensions, it is also termed the {\it quantum spin Hall state} (QSH). This state is distinct from ordinary insulators by a nonzero value of a $Z_2$ invariant.\cite{Kane:2005a,Kane:2005b} In three dimensions, there are four $Z_2$ invariants characterizing either a strong topological
insulator, a weak topological insulator, or a trivial insulator.\cite{Kane:2007,Balents:2007,Roy:prb09} The $Z_2$ invariants can be obtained via knowledge of the single-particle wave functions alone.

The key to experimental realizations of TIs (at least so far) is strong intrinsic spin-orbit interaction originating from relativistic effects. The topologically nontrivial behavior in these systems is stabilized by a strong spin-orbit coupling which leads to a ``band inversion".\cite{Bernevig:sci06,Bernevig:prl06,Murakami:prb07}  While the experimental search for the TIs in real materials with strong spin-orbit coupling is still under way with a number of examples found to date,\cite{Konig:sci07,Roth:sci09,Hsieh:nat08,Hsieh:sci09,Xia:np09,Chen:sci09,Hsieh:nat09,Hsieh:prl09} the current theoretical research in TIs is quite diverse. On the one hand, there have been intensive first-principle studies to identify potential candidate materials for TIs.\cite{Zhang:2009,Zhang:2010a,Zhang:2010b,Hasan:2010a,Hasan:2010b}
On the other hand, the study of TIs in the presence of disorder~\cite{Shen:2009,Beenakker:2009} and interplay of spin-orbit coupling and electron-electron interaction~\cite{Pesin:2010,Hur:2010} have been carried out. New exotic phases have been proposed, such as a {\it topological Mott insulator},\cite{Pesin:2010} which has a gapped charge sector but gapless spinon excitations on the boundary.

In the present paper, we focus on yet another class of systems in which the topologically nontrivial nature of the wave functions is a result of spontaneously broken symmetry in an interacting system.\cite{Raghu:2008,Yi:2009,sun:2009} These {\it interaction-driven topological insulators} possess conventional order parameters and the topological order is locked to those. Microscopically, the topological phases are described by the spontaneous generation of (spin) currents, a popular theoretical idea which has been used in many variants for describing the pseudogap phase of the cuprates.\cite{Affleck:prb88,Wen:prl96,Varma:prl99,Chakravarty:prb01}
However, in contrast to these cuprate models defined on the square lattice, a gap can be opened over the whole Brillouin zone in certain other lattices\cite{Nagaoso:2000,Raghu:2008,Yi:2009,sun:2009} allowing one to characterize the phase by a topological invariant. For example, in Ref.~[\onlinecite{Nagaoso:2000}] a double-exchange ferromagnet has been studied on the kagome lattice and the ground state has been described as a chiral spin state with a finite Chern number. Later, Raghu {\it et al.} studied an extended Hubbard model on the honeycomb lattice and showed that both a quantum anomalous Hall phase and a quantum spin Hall phase can be generated dynamically.\cite{Raghu:2008} A similar idea has also been used to obtain a three-dimensional example of an interaction-driven topological insulator on the diamond lattice.\cite{Yi:2009}

In our paper, we focus on spinless and spinful extended Hubbard models with repulsive interactions on the kagome and decorated honeycomb lattice. Interacting electrons on the kagome lattice provide a model system where ferromagnetism can be rigorously shown for certain parameters ({\it flat-band}\cite{Mielke:1992,Tasaki:1998} and {\it kinetic}\cite{Pollmann:2008} ferromagnetism). Furthermore, the Mott transition in the standard Hubbard model defined on this lattice has been studied.\cite{Hirokazu:2006} In addition to these examples, a great deal of the theoretical work on Hubbard and extended Hubbard models has focused on the case of half-filling where the low-energy degrees of freedom are described by a frustrated quantum spin model.\cite{Hermele:prb08,Wang:prb06,Singh:prb07,Evenbly:prl10} These studies are motivated in part by the recent discovery that herbertsmithite, a spin-1/2 kagome antiferromagnet, might support a spin liquid ground state.\cite{Helton:prl07,Imai:prl08} Another system where the physics of interacting electrons on the kagome lattice might be important is Na$_x$CoO$_2$ where the orbital degrees of freedom give rise to four interpenetrating kagome systems.\cite{Koshibae:2003,Indergand:2005}

The decorated honeycomb lattice can be viewed as an interpolating lattice between honeycomb and kagome. While there are few known examples of this lattice in nature,\cite{Zheng} the exact ground states of the Kitaev model on this lattice have been found by Yao $et. al$~\cite{Yao:2007} and other higher symmetry spin models have been studied as well.\cite{Yang:prb10}  Yao $et. al$~\cite{Yao:2007} has shown that the exact ground state of the Kitaev model on this lattice is a chiral spin liquid that spontaneously breaks time-reversal symmetry. There are two topologically distinct chiral spin liquid phases: (i) a topologically nontrivial phase with odd Chern number and non-abelian vortex excitations and (ii) a topologically-trivial phase with even Chern number and abelian vortex excitations. In our previous work,~\cite{Ruegg:prb10} we have found that this lattice also supports a TI phase in the presence of spin-orbit coupling at various filling fractions, and we established a connection between the topologically nontrivial chiral spin liquid state of the Kitaev model (appropriate for strongly interacting electrons with spin-orbit coupling) and the ground state of $Z_2$ topological band insulators (studied in the noninteracting limit).

Both the kagome and decorated honeycomb lattices support a TI in a single-orbital tight-binding model with spin-orbit coupling.\cite{Guo:2009,Ruegg:prb10} In this paper we show that a TI (quantum anomalous Hall state for the spinless case) can also be interaction-driven on both lattices. We focus on filling fractions which either involve a pair of Dirac points ($1/3$ filling in the kagome system) or a quadratic band crossing point ($2/3$ filling in the kagome and $1/2$ filling in the decorated honeycomb system) in the noninteracting tight-binding model.\cite{Kargarian:prb10} Using a Hartree-Fock mean-field approach, we discuss various possible symmetry broken states, present the phase diagrams and highlight the competition between various states. We find pronounced differences for different filling fractions. In particular, a topologically nontrivial phase is the leading instability at $2/3$ filling on the kagome lattice and 1/2 filling on the decorated honeycomb lattice. On the other hand, to stabilize a topologically nontrivial phase at $1/3$ filling on the kagome lattice, some fine tuning of the interaction parameters is required. We also point out that the kagome and decorated honeycomb lattices provide examples where topological phases can emerge solely due to a complex nearest neighbor hopping, in contrast to the honeycomb or diamond lattice in which a complex second neighbor hopping is required, at least within a single band model.

The paper is organized as follows. In Sec.~II, we introduce spinless and spinful extended Hubbard models on the kagome and decorated honeycomb lattices, and review the tight-binding band structures and Hartree-Fock mean-field approach for the implementation of numerical calculations. In Sec.~III and IV, we discuss several symmetry-breaking candidate phases and present phase diagrams of spinless and spinful extended Hubbard models at $1/3$ and $2/3$ filling fractions. We find the topologically nontrivial phases can be stabilized under suitable circumstances. Comparisons are also made to related work. Then, in Sec.~V we briefly discuss the spinless extended Hubbard model on the decorated honeycomb lattice. Finally, we present our conclusions and summary in Sec.~VI.

\section{Models and Methods}
We first introduce the models which will be studied later by means of the Hartree-Fock approximation. We consider both spinless (spin-polarized) and spinful interacting fermions in a single-orbital Hamiltonian on the kagome and the decorated honeycomb lattice.

\subsection{Extended Hubbard models}
The lattice model under consideration for spinless (spin-polarized) fermions takes the form
\begin{eqnarray}
&&H_{\rm spinless}=-t\sum_{\langle i,j \rangle}c_{i}^{\dag }c_{j }+V_{1}\sum_{\langle i,j \rangle }
n_{i }n_{j }\nonumber\\
&&+V_{2}\!\!\sum_{\langle\!\langle i,j \rangle\! \rangle }n_{i }n_{j }+V_{3}\!\!\sum_{\langle\! \langle\!\langle i,j \rangle\! \rangle\! \rangle }n_{i }n_{j }.
\label{eq:H_spinless}
\end{eqnarray}
Here, $c_i^{(\dag)}$ annihilates (creates) a spinless fermion on site $i$ and $n_i=c_i^{\dag}c_i^{}$ is the fermion density operator on site $i$. The sums run over nearest-neighbor $\langle i,j\rangle$, second-neighbor $\langle\!\langle i,j\rangle\!\rangle$, or third-neighbor bonds $\langle\!\langle\!\langle i,j\rangle\!\rangle\!\rangle$. The hopping amplitude is denoted by $t$ and the parameters $V_1$, $V_2$, and $V_3$ quantify the nearest-neighbor, second-neighbor and third-neighbor repulsion, respectively. For most parts of our work we set $V_3=0$. However, as we show later, a small but finite $V_3$ is necessary to stabilize a topologically non-trivial insulator for $1/3$ filling fraction on the kagome lattice.

The model for spinful fermions includes an additional on-site repulsive  interaction $U$. The Hamiltonian reads
\begin{eqnarray}
&&H_{\rm spinful}=-t\sum\limits_{\langle i,j \rangle}c_{i\sigma}^{\dag }c_{j \sigma}+U\sum\limits_{i}n_{i\uparrow }n_{i\downarrow }\nonumber\\
&&+V_{1}\sum_{\langle i,j \rangle }n_{i }n_{j }
+V_{2}\!\!\sum_{\langle\!\langle i,j \rangle\! \rangle }n_{i }n_{j }+V_{3}\!\!\sum_{\langle\! \langle\!\langle i,j \rangle\! \rangle\! \rangle }n_{i }n_{j }.
\label{eq:H_spinful}
\end{eqnarray}
Here, $c_{i\sigma}^{(\dag)}$ annihilates (creates) a fermion on site $i$ with spin $\sigma=\uparrow, \downarrow$, $n_{i\sigma}=c_{i\sigma}^{\dag}c_{i\sigma}^{}$ and $n_i=\sum_{\sigma}n_{i\sigma}$. The summing convention and the meaning of the parameters $V_1$, $V_2$, and $V_3$ are the same as for the spinless model.
\subsection{Kagome and decorated honeycomb lattice}
The models in Eqs.~\eqref{eq:H_spinless} and \eqref{eq:H_spinful} have been studied on the kagome and the decorated honeycomb lattice in the non-interacting limit.\cite{Guo:2009,Ruegg:prb10} A section of the kagome lattice is shown in Fig.~\ref{fig:kagome_flux_pattern} and a section of the decorated honeycomb lattice is shown in Fig.~\ref{fig:DH_QAH_CDW_structure}. Both lattices share an underlying triangular lattice and we choose the unit cell vectors to be
\begin{equation}
{\boldsymbol  a}_{1}=(a,0)\quad {\rm and}\quad {\boldsymbol  a}_{2}=(\frac{a}{2},\frac{\sqrt{3}}{2}a),
\end{equation}
where $a$ is their length. The kagome lattice has three sites in the unit cell whereas the decorated honeycomb lattice has six. The reciprocal lattice vectors are given by
\begin{equation}
{\boldsymbol b}_{1}=\frac{2\pi}{a} (1,\frac{-1}{\sqrt{3}})\quad {\rm and} \quad
{\boldsymbol b}_{2}=\frac{2\pi}{a} (0,\frac{2}{\sqrt{3}}).
\end{equation}
The first Brillouin zone forms a hexagon in momentum space for both lattices, similar to the honeycomb lattice which also shares the underlying triangular lattice.
\subsubsection{Tight-binding band structure on kagome lattice}
The noninteracting energy dispersion for a nearest-neighbor tight-binding model [first term in Eq.~\eqref{eq:H_spinless}] can be obtained analytically. On the kagome lattice, three bands are found with the following dispersion relation:
\begin{equation}
\epsilon_{1}(\boldsymbol{k})=-t-t A_{\boldsymbol{k}},\quad\epsilon_{2}(\boldsymbol{k})=-t+t A_{\boldsymbol{k}},\quad \epsilon_{3}(\boldsymbol{k})=2t.
\label{eq:kagome_disp}
\end{equation}
In Eq.~\eqref{eq:kagome_disp} we have defined
\begin{equation}
A_{\boldsymbol{k}}=\sqrt{3+2\cos k_{1}+2\cos k_{2}+2\cos (k_{1}-k_{2})},
\label{eq:Ak}
\end{equation}
where $k_1={\boldsymbol a}_1\cdot {\boldsymbol k}$ and $k_2={\boldsymbol a}_2\cdot{\boldsymbol k}$. There are two dispersing bands ($n=1$ and 2) and a flat band ($n=3$). At filling fraction $f=1/3$, the two dispersing bands touch at two inequivalent Dirac points located at corners of the Brillouin zone
\begin{equation}
\boldsymbol{K}_{\pm}=\pm(\boldsymbol{b}_{1}-\boldsymbol{b}_{2})/3.
\end{equation}
At filling fraction $f=2/3$, the second band touches the flat band at the $\boldsymbol{\Gamma}$ point [${\boldsymbol k}=(0,0)]$. This is a quadratic band crossing point (QBCP).\cite{sun:2009} Upon inclusion of an intrinsic spin-orbit coupling (modeled by a spin-dependent imaginary second-neighbor hopping) one finds that a gap is opened both at the Dirac points ($f=1/3$) and the QBCP ($f=2/3$).\cite{Guo:2009} The resulting insulating state
at $f=1/3$ and $f=2/3$ is a $Z_2$ topological insulator with time-reversal symmetry protected edge states.\cite{Guo:2009} In the following sections, we explore the possibility of dynamically generating a topological insulator phase from interactions and study its competition with other broken-symmetry phases. We therefore focus on $f=1/3$ and $f=2/3$ in this paper.

\subsubsection{Tight-binding band structure on decorated honeycomb lattice}
Diagonalization of the noninteracting tight-binding model on the decorated honeycomb lattice gives the following six bands:
\begin{subequations}
\begin{align}
\varepsilon_1({\boldsymbol k})&=-\frac{t}{2}-\sqrt{\frac{9}{4}t^2+t'^2+tt' A_{\boldsymbol k}},\\
\varepsilon_2({\boldsymbol k})&=-\frac{t}{2}-\sqrt{\frac{9}{4}t^2+t'^2-tt' A_{\boldsymbol k}},\\
\varepsilon_3({\boldsymbol k})&=t-t',\\
\varepsilon_4({\boldsymbol k})&=-\frac{t}{2}+\sqrt{\frac{9}{4}t^2+t'^2-tt' A_{\boldsymbol k}},\\
\varepsilon_5({\boldsymbol k})&=-\frac{t}{2}+\sqrt{\frac{9}{4}t^2+t'^2+tt' A_{\boldsymbol k}},\\
\varepsilon_6({\boldsymbol k})&=t+t',
\end{align}
\end{subequations}
and $A_{\boldsymbol k}$ is defined in Eq.~\eqref{eq:Ak}. Here, we have introduced independent hopping amplitudes for hopping within a triangle ($t$) and between triangles ($t'$).\cite{Ruegg:prb10} There are two flat bands ($n=3,6$) and four dispersing bands ($n=1,2,4,5$). For filling fractions $1/6$ and $2/3$, there are Dirac points located at $\boldsymbol{K}_{\pm}$ in the momentum space. There are also two quadratic band touching points at ${\boldsymbol k}=(0,0)$. The lower QBCP appears at $f=1/2$ if $t'<3t/2$ and at $f=1/3$ if $t'>3t/2$. The upper QBCP appears at $f=5/6$. In the presence of a spin-orbit coupling, TI phases are found at various filling fractions.\cite{Ruegg:prb10} In this paper we set $t=t'$ and solely focus on $f=1/2$. Half filling is of particular interest because a topological connection between the chiral spin liquid states recently found in the Kitaev model\cite{Yao:2007} and the $Z_2$ topological band insulator has been established.\cite{Ruegg:prb10}

\subsection{Hartree-Fock mean-field approximation}
We use the standard Hartree-Fock mean-field approach to decouple the interaction terms in Eqs.~\eqref{eq:H_spinless} and \eqref{eq:H_spinful}. In contrast to comparable studies on the honeycomb lattice\cite{Raghu:2008,Weeks:2010} and the diamond lattice,\cite{Yi:2009} we treat the Hartree and Fock terms on equal footing in all phases.

\subsubsection{Hartree-Fock approximation in the spinless models}
For spinless fermions, we decouple the interaction both in the direct and the exchange channel:
\begin{eqnarray}
n_{i}n_{j}&\approx& n_{i}\langle n_{j}\rangle+\langle n_{i}\rangle n_{j}-\langle n_{i}\rangle\langle
n_{j}\rangle\nonumber\\
&&-c_{i}^{\dag}c_{j}\langle c_{j}^{\dag}c_{i}\rangle-\langle c_{i}^{\dag}c_{j}\rangle c_{j}^{\dag}c_
{i}
+\langle c_{i}^{\dag}c_{j}\rangle\langle c_{j}^{\dag}c_{i}\rangle.
\label{eq:HFspinless}
\end{eqnarray}
This procedure yields a mean-field Hamiltonian which is bilinear in the fermionic operators and can be diagonalized. In the following, we focus on uniform phases which are characterized by a (possibly enlarged) unit cell. We work in the canonical ensemble with a fixed number of
electrons $N_e$. The free energy at temperature $k_BT=\beta^{-1}$ is given by
\begin{eqnarray}
F&=&-k_{B}T\sum\limits_{\boldsymbol{k},n}\log\left[1+e^{\beta (E_{\boldsymbol{k}n}-\mu)}\right]
+\mu N_e\nonumber\\
&+&V_1\sum_{\langle i,j \rangle}\left(\langle c_{i}^{\dag}c_{j}\rangle\langle c_{j}^{\dag}c_{i}
\rangle-\langle n_{i}\rangle\langle n_{j}\rangle\right)\nonumber\\
&+&(V_2, V_3){\rm -terms},
\label{eq:free}
\end{eqnarray}
where the chemical potential $\mu=\mu(T,N_e)$. The terms in the second and third line of Eq.~\eqref{eq:free} arise from $\langle n_{i}\rangle\langle n_j\rangle$ and $\langle c_{i}^{\dag}c_{j}\rangle \langle c_{j}^{\dag}c_{i}\rangle$ in the decoupling Eq.~\eqref{eq:HFspinless} and are not included in the single-particle energies $E_{\boldsymbol{k}n}$. The most general self-consistency (mean-field) equations are
\begin{equation}
\frac{\partial F}{\partial \langle n_{i}\rangle}=\frac{\partial F}{\partial \langle c_{i}^{\dag}c_{j}
\rangle}=\frac{\partial F}{\partial \langle c_{j}^{\dag}c_{i}\rangle}=0.
\label{eq:sce}
\end{equation}
In the following sections, we discuss various solutions of these equations.

\subsubsection{Hartree-Fock approximation in the spinful models}
For spinful fermions, we decouple the on-site interaction according to
\begin{eqnarray}
&n_{i\uparrow}n_{i\downarrow}\approx n_{i\uparrow}\langle n_{i\downarrow}\rangle+\langle n_{i\uparrow}\rangle n_{i\downarrow}-\langle n_{i\uparrow}\rangle\langle n_{i\downarrow}\rangle\nonumber\\
&-c_{i\uparrow}^{\dag}c_{i\downarrow}\langle c_{i\downarrow}^{\dag}c_{i\uparrow}\rangle-\langle c_{i\uparrow}^{\dag}c_{i\downarrow}\rangle c_{i\downarrow}^{\dag}c_{i\uparrow}+\langle c_{i\uparrow}^{\dag}c_{i\downarrow}\rangle\langle c_{i\downarrow}^{\dag}c_{i\uparrow}
\rangle.
\end{eqnarray}
We assume the mean-field solutions are described by a co-linear spin alignment and therefore, without loss of generality, we set $\langle c_{i\uparrow}^{\dag}c_{i\downarrow}\rangle=\langle c_{i\downarrow}^{\dag}c_{i\uparrow}\rangle=0$ in what follows. For the model on the kagome lattice with $V_1=V_2=V_3=0$, and at filling fractions $f=1/3$ and $f=2/3$, we have explicitly checked that with {\it all} the terms (including $\langle c_{i\uparrow}^{\dag}c_{i\downarrow}\rangle$) all our self-consistent solutions indeed have a co-linear spin alignment. We expect this property will persist also for finite further neighbor interactions. However, the $\langle c_{i\uparrow}^{\dag}c_{i\downarrow}\rangle$ term has to be kept if one works at half filling on the kagome lattice where at the mean-field level a coplanar $120^{\circ}$ antiferromagnetic state arises in the large $U$ limit. The same antiferromagnetic state has also been found on the triangular lattice.\cite{Huse:1988,Pierre:1994,Avinash:2005}

The further-neighbor interaction is decoupled in a similar way:
\begin{eqnarray}
n_{i}n_{j}&\approx& n_{i}\langle n_{j}\rangle+\langle n_{i}\rangle n_{j}-\langle n_{i}\rangle\langle n_{j}\rangle-\sum_{\alpha\beta}\left(c_{i\alpha}^{\dag}c_{j\beta}\langle c_{j\beta}^{\dag}c_{i\alpha}\rangle\right.\nonumber\\
&&+\left.\langle c_{i\alpha}^{\dag}c_{j\beta}\rangle c_{j\beta}^{\dag}c_{i\alpha}-\langle c_{i\alpha}^{\dag}c_{j\beta}\rangle\langle c_{j\beta}^{\dag}c_{i\alpha}\rangle\right).
\end{eqnarray}
Again, as mentioned above, we set $\langle c_{i\alpha}^{\dag}c_{j\beta}\rangle=0$
for $\alpha\neq\beta$ which is justified if the spin alignment is co-linear in the physical solutions. The structure of the free energy and the self-consistency equations are similar to Eqs.~\eqref{eq:free} and \eqref{eq:sce} for the spinless models.

\section{Spinless fermions on kagome lattice}
In this section we discuss the zero temperature Hartree-Fock mean-field phase diagrams at filling fractions $f=1/3$ and $f=2/3$ for the spinless model on the kagome lattice. We first introduce the candidate phases and then show the $V_1$-$V_2$ phase diagrams with and without a finite $V_3$ for the two special filling fractions. Because at $f=1/3$ there are Dirac points involved, and at $f=2/3$ there is a QBCP, the phase diagrams look rather different for these two cases.

\subsection{Candidate phases}
Let us now introduce possible candidate phases for the spinless model.
Besides the topologically non-trivial quantum anomalous Hall (QAH) phase we also take into account possible charge density wave (CDW) patterns.

\subsubsection{Quantum anomalous Hall phase}
\begin{figure}[tbp]
\centering
\includegraphics[width=0.5\linewidth]{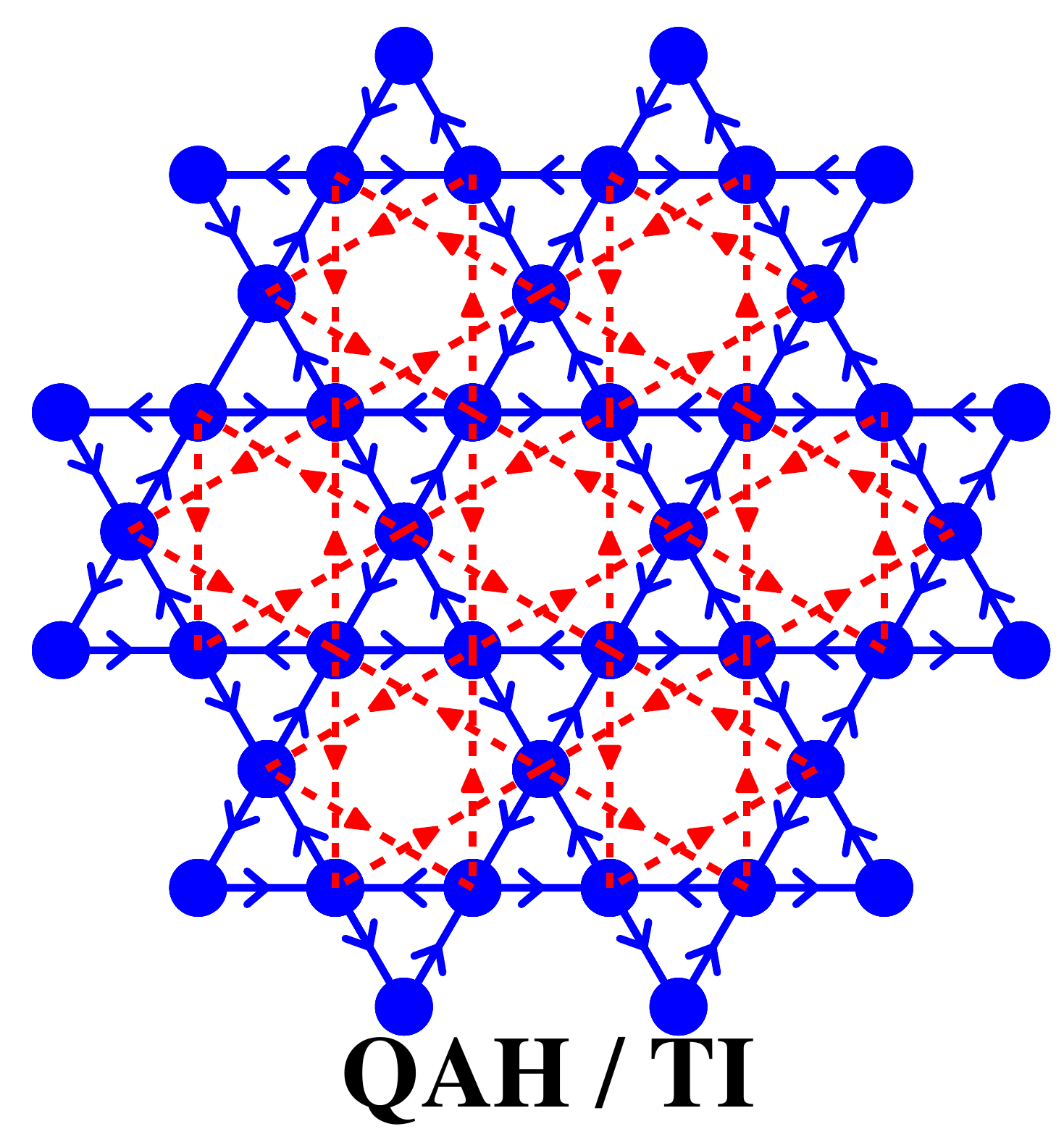}
\caption{(Color online) The spinless flux pattern developed by nearest and second nearest interactions that preserves lattice symmetry but spontaneously breaks time-reversal symmetry on the kagome lattice. Charges are uniform on all sites. The blue solid(red dash) line represents a (second)nearest neighbor complex hopping. For the spinful case, two copies of the same(opposite) flux patterns for spin-up and spin-down fermions form the quantum anomalous Hall (topological insulator) state. }
\label{fig:kagome_flux_pattern}
\end{figure}

A complex Fock term in Eq.~\eqref{eq:HFspinless} breaks time-reversal symmetry and can give rise to a topological phase characterized by a non-vanishing Chern number\cite{Raghu:2008} even though there is no external magnetic field. In the present case, the total flux through the unit cell must be zero (this follows from periodic boundary conditions on the unit cell). However, there are finite fluxes through the elementary loops and the system shows an integer quantum Hall effect. This is in full analogy to Haldane's model on the honeycomb lattice.\cite{Haldane:1988} Such a state of matter is called a quantum anomalous Hall phase and a schematic illustration of its microscopic current pattern on the kagome lattice with finite $V_1$ and $V_2$ is shown in Fig.~\ref{fig:kagome_flux_pattern}.

The QAH phase preserves the translational symmetry of the noninteracting model but breaks time-reversal symmetry. A solution of the self-consistency Eqs.~\eqref{eq:sce} is obtained by assuming a uniform charge distribution and introducing complex bond expectation values. For nearest neighbor bonds we make the following {\it ansatz}:
\begin{equation}
\langle c_{i}^{\dag}c_{j}\rangle=\chi\exp({i\varphi_{ij}})=\chi_1+i\chi_2.
\end{equation}
A similar {\it ansatz} is also made for second-nearest neighbor bonds:
\begin{equation}
\langle c_{i}^{\dag}c_{j}\rangle=\chi^{\prime}\exp({i\varphi_{ij}^{\prime}})=\chi_1'+i\chi_2'.
\end{equation}
There is a gauge freedom in choosing the phase factors $\varphi_{ij}$ and $\varphi_{ij}^{\prime}$ because only the inclosed fluxes through elementary loops are gauge invariant. We choose a uniform gauge $\varphi_{ij}^{(\prime)}=\pm\varphi^{(\prime)}$ where the sign is fixed according to the dictions of the arrows in Fig.~\ref{fig:kagome_flux_pattern}.
We stress that on the kagome lattice a complex
nearest-neighbor hopping can already stabilize a topologically non-trivial phase showing an integer quantum Hall effect. This possibility has been explored in a model of a ferromagnet with spin anisotropy.\cite{Nagaoso:2000} Therefore, in contrast to the honeycomb\cite{Raghu:2008,Weeks:2010} and the diamond lattice,\cite{Yi:2009} the nearest-neighbor interaction $V_1$ alone can in principle generate a QAH phase if the time reversal symmetry is spontaneously broken. Indeed, we show below that at $f=2/3$ this is the case. However, at $f=1/3$ we find it essential to have a finite $V_2$ and small $V_3$ in order to  stabilize the QAH state.

\subsubsection{Charge density waves}

An effective way to lower the potential energy is to develop an inhomogeneous charge distribution. In the atomic limit $t=0$ and in the absence of further neighbor interactions, $V_2=V_3=0$, there is a macroscopically degenerate set of charge configurations which minimize the energy. At $f=1/3$ ($f=2/3$) these configurations obey the ``one particle (hole) per triangle"- rule. A finite $t$ lifts the degeneracy and in the limit $t/V_1\ll 1$ the system is effectively described by a hardcore dimer model on the honeycomb lattice.\cite{Pollmann:2008,nishimoto:2010,OBrien:2010} Its ground state is the ``plaquette" phase with resonating plaquettes and a periodicity which triples the unit cell.\cite{Moessner:2001} Physically, it is the ring exchange of order $|t|^3/V_1^2$ which stabilizes the
plaquette phase. We also note that in the limit $t/V_1\ll 1$ the system becomes particle-hole symmetric. This property is clearly lost for small to intermediate interactions, see below.

\begin{figure}[tbp]
\centering
\includegraphics[width=\linewidth]{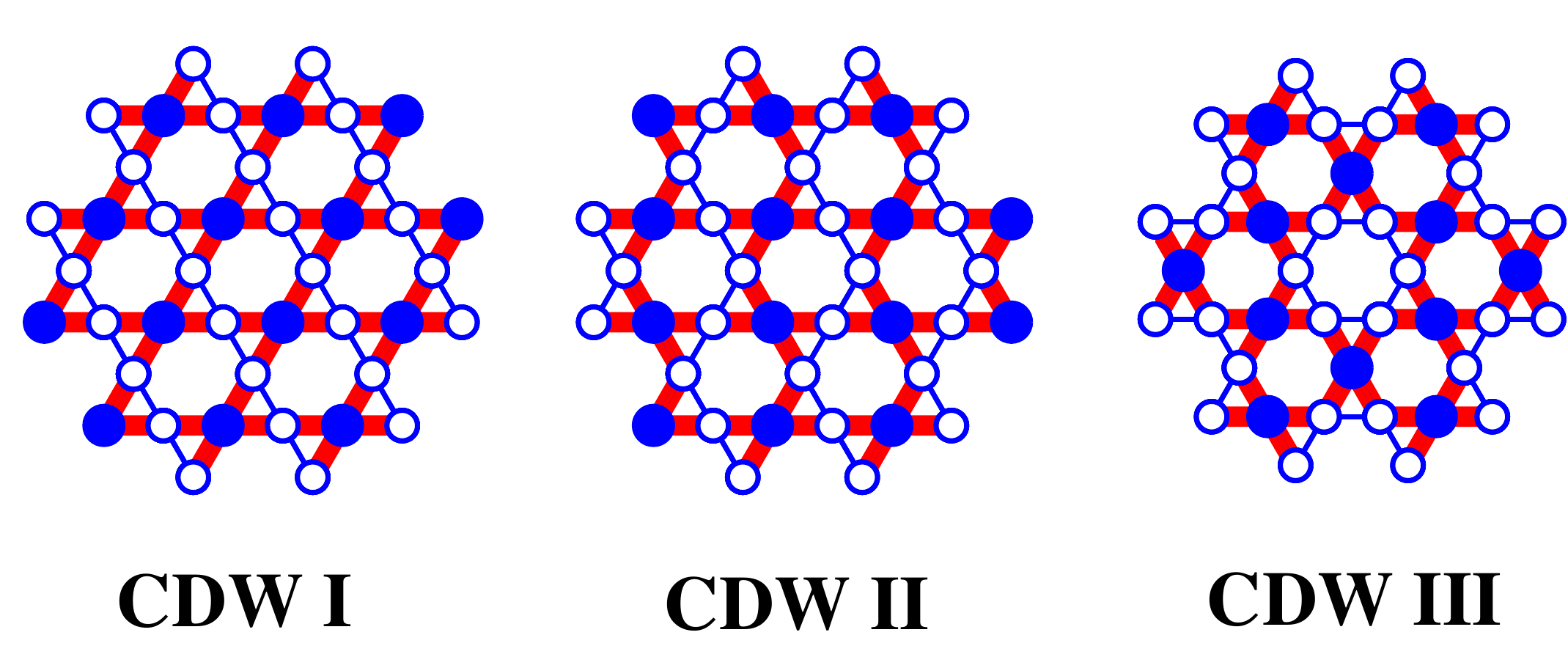}
\caption{(Color online) Three different charge density wave patterns on the kagome lattice studied in this paper. They are characterized by the  wave vectors ${\boldsymbol q}_{\rm I}=(0,0)$, ${\boldsymbol q}_{\rm II}={\boldsymbol b}_2/2$ and $\boldsymbol{ q}_{\rm III}=({\boldsymbol b}_1-{\boldsymbol b}_2)/3$. Blue sites stand for the fermion-rich (poor) sites at 1/3 (2/3) filling fraction and white sites for the fermion-poor (rich) sites at 1/3 (2/3) filling fraction. The bond expectation values oscillate in the real space as well. We distinguish strong and weak bonds by thick and thin lines. For simplicity, we do not show the second neighbor bonds.}
\label{fig:kagome_CDW_pattern}
\end{figure}
Further neighbor interactions $V_2, V_3>0$ lift the degeneracy of the charge configurations in the atomic limit. This fact complicates a mapping to an effective dimer model for finite $t$. In the following we study the mean-field solutions of a limited number of different classical charge distributions. Specifically, we consider three different charge density wave (CDW) patterns which we denote by I, II and III; see Fig.~\ref{fig:kagome_CDW_pattern}. They were introduced in Ref.~[\onlinecite{nishimoto:2010}] in order to numerically study the role of the ring exchange. For us it is important to realize that $V_2>0$ favors CDW~I as compared to CDW~II and III in the atomic limit. On the other hand, a third neighbor interaction $V_3>0$ favors CDW~III over CDW~I and CDW~II. While the unit cell of pattern I is equal to the noninteracting unit cell, the unit cell of pattern II is doubled and the one of pattern III is tripled. Note that CDW~III can be viewed as the classical charge distribution which corresponds to the plaquette phase of the effective dimer model in the limit $t/V_1\ll 1$.

For the CDW~I, the wave vector specifying its periodicity is ${\boldsymbol q}_{\rm I}=(0,0)$ and the densities on the three inequivalent sites of the noninteracting unit cell are given by
\begin{eqnarray}
\langle n_1(\boldsymbol{r}_{nm})\rangle_{I}&=&f+\rho_1,\nonumber\\
\langle n_2(\boldsymbol{r}_{nm})\rangle_{I}&=&f+\rho_2,\nonumber\\
\langle n_3(\boldsymbol{r}_{nm})\rangle_{I}&=&f+\rho_3,
\label{eq:nrI}
\end{eqnarray}
where $\boldsymbol{r}_{nm}=n\boldsymbol{a}_1+m\boldsymbol{a}_2$ with $(n,m)\in \mathbb{Z}\times\mathbb{Z}$, $f$ is the filling fraction and $\rho_1+\rho_2+\rho_3=0$. Similarly, the densities in the CDW~II configuration can be written as
\begin{eqnarray}
\langle n_1(\boldsymbol{r}_{nm})\rangle_{II}&=f+\rho_1 \cos({\boldsymbol{r}_{nm}\cdot
\boldsymbol{q}_{II}}),\nonumber\\
\langle n_2(\boldsymbol{r}_{nm})\rangle_{II}&=f+\rho_2 \cos({\boldsymbol{r}_{nm}\cdot
\boldsymbol{q}_{II}}),\nonumber\\
\langle n_3(\boldsymbol{r}_{nm})\rangle_{II}&=f+\rho_3 \cos({\boldsymbol{r}_{nm}\cdot
\boldsymbol{q}_{II}}),
\label{eq:nrII}
\end{eqnarray}
where we have introduced the wave vector ${\boldsymbol q}_{\rm II}={\boldsymbol b}_2/2$. In our mean-field calculations we find that mirror symmetric charge configurations are always favored. Such configurations are obtained by setting $\rho_1=2\rho$ and $\rho_2=\rho_3=-\rho$ (or cyclically permuted) in Eqs.~\eqref{eq:nrI} and \eqref{eq:nrII}. CDW~I and II both break the six-fold rotations symmetry ($C_6$) of the kagome lattice; CDW~I breaks it down to $C_2$, while CDW~II breaks it down even futher.  In both cases, there are three different possibilities to choose a mirror symmetry plane. The CDW order parameter therefore has an additional $Z_3$ freedom.

Another phase with a mirror symmetric configuration of the densities in the non-interacting unit cell is described by the CDW~III pattern:

\begin{eqnarray}
\langle n_1(\boldsymbol{r}_{nm})\rangle_{III}&=&f+2\rho\cos\left[{\boldsymbol{r}_{nm}\cdot
\boldsymbol{q}_{III}}+s\frac{2\pi}{3}\right],\nonumber\\
\langle n_2(\boldsymbol{r}_{nm})\rangle_{III}&=&f+2\rho\cos\left[{\boldsymbol{r}_{nm}\cdot
\boldsymbol{q}_{III}}+(s-1)\frac{2\pi}{3}\right],\nonumber\\
\langle n_3(\boldsymbol{r}_{nm})\rangle_{III}&=&f+2\rho\cos\left[{\boldsymbol{r}_{nm}\cdot
\boldsymbol{q}_{III}}+(s+1)\frac{2\pi}{3}\right],
\label{eq:nr}
\end{eqnarray}
where the wave vector is ${\boldsymbol q}_{\rm III}=({\boldsymbol b}_1-{\boldsymbol b}_2)/3$. We have introduced the parameter $s=0,1,2$ which characterizes the $Z_3$ freedom in the CDW~III. Changing the value of $s$ results in a shift of the pattern as a whole either by $\boldsymbol{a}_1$ or $\boldsymbol{a}_2$.

CDW~I does not break the original translation symmetry and it can alternatively be viewed as a nematic phase.\cite{sun:2009} The direction associated with the nematic order is given by
\begin{equation}
\boldsymbol{e}=(Q_x,Q_y)/\sqrt{Q_x^2+Q_y^2},
\label{eq:e}
\end{equation}
where the components are obtained from the charge and bond order
\begin{eqnarray}
&&Q_x=\frac{\langle n_1\rangle-\langle n_2\rangle}{\sqrt{3}}\pm2\frac{\langle c^{\dag}
_2c_3\rangle-\langle c^{\dag}_1c_3\rangle}{\sqrt{3}},\\
&&Q_y=\frac{\langle n_1\rangle\!+\!\langle n_2\rangle\!-\!2\langle n_3\rangle}{3}\pm 2\frac{\langle
c^{\dag}_1c_3\rangle\!+\!\langle c^{\dag}_2c_3\rangle\!-\!2\langle c^{\dag}_1c_2\rangle}{3},
\nonumber
\end{eqnarray}
where the ``$-$" sign refers to the case of 1/3 filling presently being considered, and the ``$+$" sign refers to the case of 2/3 filling involving a quadratic band touching point.
This definition of the nematic order parameter is in agreement with the definition given in Eq.~(4) of Ref.~[\onlinecite{sun:2009}] for the case of 2/3 filling.

In our study, we assume that the real hopping expectation values $\chi_{ij}=\langle c_i^{\dag}c_j\rangle$ obey the same symmetry as the charge distribution. In Fig.~\ref{fig:kagome_CDW_pattern}, the weak and strong nearest-neighbor bonds are schematically shown. We find that taking into account this {\it bond order} can significantly lower the energy as compared to the case where only the Hartree term is kept. For the mirror symmetric solutions, the unit vector in Eq.~\eqref{eq:e} assumes only three different directions:
\begin{equation}
\label{eq:en}
\boldsymbol{e}_1=(\sqrt{3},1)/2,\quad \boldsymbol{e}_2=(-\sqrt{3},1)/2,\quad \boldsymbol{e}_3=(0,-1).
\end{equation}
These unit vectors will also appear in the low energy description of the CDW phases.

\subsubsection{Other phases}
Let us now briefly comment on other possible phases which are
not stabilized in the present models. Dimerized and trimerized
phases were considered in Ref.~[\onlinecite{Guo:2009}] as perturbations
to the TI phase in the noninteracting limit. We find that for a self-consistency solution with dimerized or trimerized bonds, it is crucial to take into account the charge ordering which results from the bond order. However, our numerical results suggest that the charge density wave patterns shown in Fig.~\ref{fig:kagome_CDW_pattern} (where the bond order has the same symmetry as the charge order) have lower energies than the dimerized or trimerized states. We also note that we do not find  a mixed QAH and CDW phase on kagome lattice, which is in contrast to the findings on the checkerboard lattice.\cite{sun:2009}

\subsection{Phase diagrams at $1/3$ filling fraction}
\label{sec:PD1o3}

\begin{figure}[tbp]
\centering
\includegraphics[width=\linewidth]{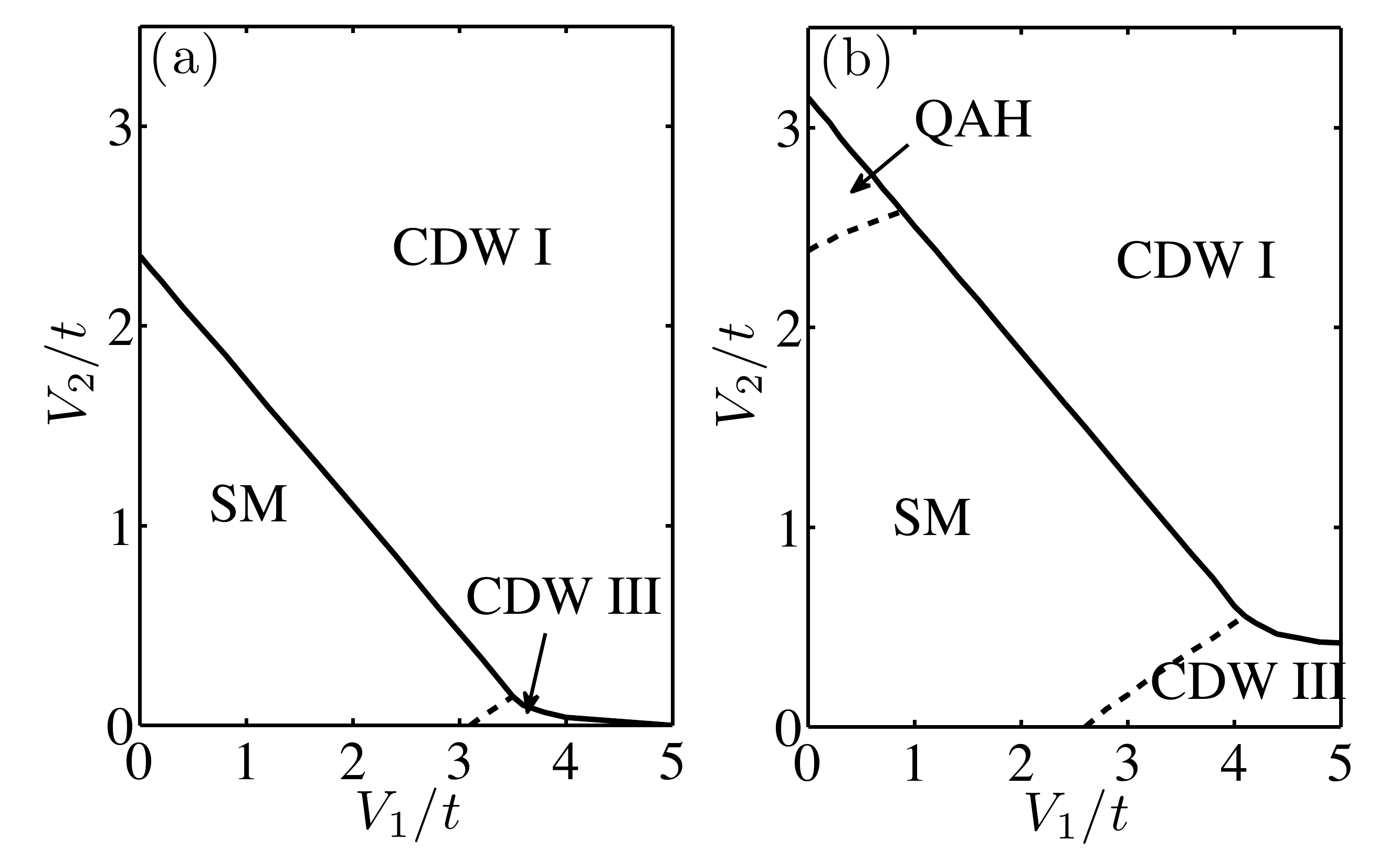}
\caption{The phase diagram of the extended Hubbard model for spinless fermions at $1/3$ filling fraction on the kagome lattice. The third neighbor interaction is (a) $V_3=0$ and (b) $V_3=0.4t$. SM denotes the semi-metallic phase with two Dirac points, QAH denotes a time-reversal symmetry broken quantum anomalous Hall phase and CDW~I and III are charge density waves with patterns shown in Fig.~\ref{fig:kagome_CDW_pattern}. Solid lines denote first and dashed lines second order transitions.}
\label{fig:kagome_phase1}
\end{figure}

Figure.~\ref{fig:kagome_phase1} shows the $V_1$-$V_2$ phase diagrams for (a) $V_3=0$ and (b) $V_3=0.4t$. At $1/3$ filling, the noninteracting Fermi ``surface" consists of a pair of Dirac points located at $\boldsymbol{K}_{\pm}$ and the density of states vanishes linearly at the Fermi energy. As in related studies,\cite{Raghu:2008,Yi:2009} our mean-field calculations yield a stable semi-metallic (SM) phase for small to intermediate interactions which can be
attributed to the absence of density of states at the Fermi level in the non-interacting limit.

\subsubsection{CDW phases and nematic order at $f=1/3$}
For large interactions, a CDW phase is stabilized. We find CDW~I for large $V_1$ and $V_2$ because both the nearest neighbor and second neighbor interaction favors CDW~I. The transition from the SM to the CDW~I is first order which is different from the situation on the honeycomb lattice.\cite{Raghu:2008}  Below we discuss this aspect in more detail. On the other hand, CDW~III is favored for small $V_2$ and large $V_1$.  The transition from the SM to CDW~III is second order.

The self-consistent CDW solutions at $f=1/3$ are always gapped (this is in contrast to $f=2/3$ where CDW phases with two nodes appear, see Sec.~\ref{sec:PD2o3}) and it is instructive to look at the corresponding low energy models. We first consider the possibility of a weak CDW~I phase and then argue that it is energetically not favored. In fact, only for a large enough order parameter, does the CDW~I solution have lower energy than the SM phase.  For simplicity, we keep only the Hartree terms. In lowest order in $\bar{V}\rho$ ($\bar{V}=V_1+V_2-2V_3$), the effective low energy Hamiltonian for the two nodes $l=\pm$ is given by
\begin{equation}
H_{\rm I}=v\sum_{\boldsymbol{k},l,\alpha,\beta}c_{\boldsymbol{k}l\alpha}^{\dag}[\boldsymbol
{\tau}\cdot(\boldsymbol{k}-l\boldsymbol{\mathcal{A}})]_{\alpha\beta}c_{\boldsymbol{k}l\beta}+
\frac{3v^2}{2\bar{V}}|\boldsymbol{\mathcal{A}}|^2,
\label{eq:heff}
\end{equation}
where the velocity is $v=\sqrt{3}ta/2$ and $\boldsymbol{\tau}=(\tau_x,\tau_y)$ are Pauli matrices in the effective ``sublattice" space. Furtherrmore, we have introduced an ``axial gauge field"\cite{Guo:2009} $\boldsymbol{\mathcal{A}}$ which can be expressed in terms of the CDW order parameter $\rho$ and the vector ${\boldsymbol e}_n$ specifying the $Z_3$ freedom:
\begin{equation}
\boldsymbol{\mathcal{A}}=-\frac{2\bar{V}}{v}\rho(\hat{n}_z\times\boldsymbol{e}_n),
\end{equation}
where the $\boldsymbol{e}_n$ are given by Eq.~\eqref{eq:en}.  This field shifts the position of the Dirac nodes with respect to their original position at $\boldsymbol{K}_{\pm}$ and consequently, the CDW~I described by Eq.~\eqref{eq:heff} has nodes. However, Eq.~\eqref{eq:heff} also includes the electron-electron interaction in the mean-field description which gives rise to the second term. This term is proportional to $v^2/\bar{V}$ and can be viewed as a mass-term for the gauge field. In other words, shifting the nodes by the vector $\boldsymbol{\mathcal{A}}$ costs an energy proportional to $|\boldsymbol{\mathcal{A}}|^2$. Therefore, it is energetically not favorable to built up a finite field $\boldsymbol{\mathcal{A}}$ and the SM phase is stable. But once $\bar{V}$ is big enough, the description in terms of Eq.~\eqref{eq:heff} breaks down. Solving the full self-consistency equations, we find a first order transition from the SM to the gapped CDW~I phase.

Let us now consider CDW~III which is stable for small $V_2$. The wave vector $\boldsymbol{q}_{\rm III}$ of pattern III connects the two inequivalent Dirac points at $\boldsymbol{K}_{\pm}$. From a weak-coupling point of view, CDW~III therefore opens a gap by coupling the two Dirac points. This can be made explicit by studying the low energy mean-field Bloch Hamiltonian. For simplicity, we set $V_2=V_3=0$ and consider only the Hartree terms. The Bloch Hamiltonian for the low energy degrees of freedom is expressed in the $4\times4$ matrix,
\begin{equation}
H_{\rm III}(\boldsymbol{k})=\begin{pmatrix}
\hat{h}^{(+)}(\boldsymbol{k}) & \hat{\Delta}\\
\hat{\Delta}^{\dag} &\hat{h}^{(-)}(\boldsymbol{k})
\end{pmatrix}.
\label{eq:HIII}
\end{equation}
Here, the Dirac Hamiltonians at $\boldsymbol{K_{\pm}}$ are given by
\begin{equation}
\hat{h}^{(\pm)}(\boldsymbol{k})=v\boldsymbol{k}\cdot\boldsymbol{\tau}.
\end{equation}
The coupling between the two Dirac cones can be brought into the following form
\begin{equation}
\hat{\Delta}=2V_1\rho
\begin{pmatrix}
1& 0\\
0 &-1
\end{pmatrix}.
\end{equation}
Equation~\eqref{eq:HIII} can be diagonalized and we find the following doubly degenerate energy bands
\begin{equation}
E_{\pm}(k)=\pm\sqrt{v^2k^2+4\rho^2V_1^2}.
\end{equation}
In particular, an arbitrarily small coupling $V_1\rho$ continuously opens a gap $4V_1|\rho|$ at the $\boldsymbol{\Gamma}$ point in the reduced Brillouin zone. (The enlarged unit cell of CDW~III moves the low-energy point from $\boldsymbol{K_\pm}$ to $\boldsymbol{\Gamma}$.)  This means that the CDW~III is a low-energy instability of the SM phase and explains why we observe a second order transition at a critical interaction strength. We also note that the low-energy theory for the CDW~III, Eq.~\eqref{eq:HIII}, carries similarities with the one found for the Kekul\'e texture on the honeycomb lattice\cite{Mudry:2007} or in the $\pi$-flux model on the square lattice.\cite{Weeks:2010} In analogy with these examples, we expect that topological defects of the CDW~III pattern in the form of a $Z_3$ vortex can give rise to interesting physics; potentially including charge fractionalization and anyon statistics.\cite{Seradjeh:2008}
 \subsubsection{Topological phase at $f=1/3$}
As shown in Fig.~\ref{fig:kagome_phase1}(b) we find that a QAH phase can be stabilized in a certain region of parameter space. Nevertheless, it requires some fine tuning of the different interaction strengths. First, we do not find a QAH solution for $V_1$ alone in the parameter space we considered (which is different from what we find at $2/3$ filling, see Sec.~\ref{sec:PD2o3}). Second, for a moderate $V_2$ there exists a self-consistent solution of Eqs.~\eqref{eq:sce} which breaks time reversal symmetry. This QAH phase is triggered by $\chi_2$ and $\chi_2^{\prime}$, the imaginary part of the nearest and second-neighbor hopping expectation values (which in general also acquire finite real parts). It turns out that for $V_3=0$, the CDW~I phase has lower energy compared to the QAH solution. However, a finite $V_3$ increases the energy of the CDW~I solution making the interaction-driven QAH phase the ground state for small $V_1$ and $V_3$ and moderate $V_2$, as shown in Fig.~\ref{fig:kagome_phase1}(b).

\subsection{Phase diagram at $2/3$ filling}
\label{sec:PD2o3}

\begin{figure}[tbp]
\centering
\includegraphics[width=\linewidth]{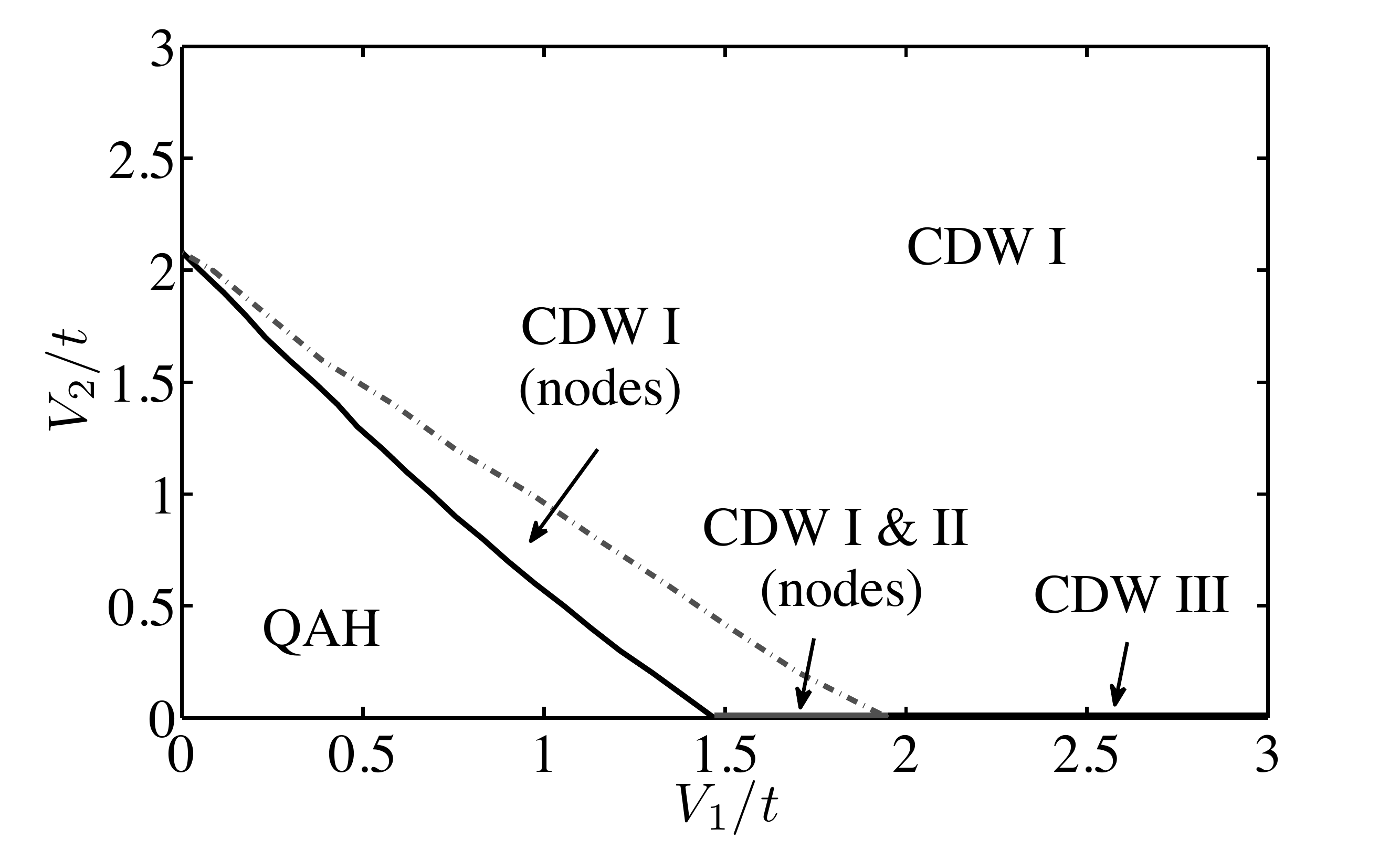}
\caption{(Color online) The phase diagram of the extended Hubbard model for spinless fermions at $2/3$ filling on the kagome lattice. The dotted line indicates where the gap opens in the CDW~I phase. For $V_2=0$, CDW~I and II coexist in the gray region and CDW~III in the black region. We set $V_3=0$.}
\label{fig:kagome_phase2}
\end{figure}

The phase diagram of the extended Hubbard model for spinless fermions at $2/3$ filling is shown in Fig~\ref{fig:kagome_phase2}. Here, we set $V_3=0$. The important difference with $1/3$ filling is that the Fermi energy in the noninteracting case lies at a QBCP between a dispersing and a flat band. As a consequence, the density of states is finite at the Fermi energy and the system is unstable to arbitrarily weak interactions.\cite{sun:2009} In particular, the semi-metallic phase does not survive even for small values of the interactions. The phase diagram for low to intermediate interactions looks therefore quite different  than the corresponding phase diagram at $1/3$ filling.
\subsubsection{Topological phase at $f=2/3$}
For small to intermediate interactions we find that the QAH phase has the lowest energy. This is in agreement with quite general arguments made about the stability of a QBCP.\cite{sun:2009} We have numerically calculated the Chern number~\cite{Fukui:2005} associated with this state and found that it is $\pm1$, indicating it is indeed a topological state displaying an integer quantum Hall effect. Note that $V_1$ alone is enough to generate the QAH phase because of the particular geometry of the kagome lattice with a triangle in the unit cell.

Although the QAH phase is the ground state in a rather large region of parameter space, its gap is exponentially small. The exponential dependence in mean-field theory can be found by analyzing the gap equation derived from an effective two band Hamiltonian describing the low-energy behavior around the QBCP. Let us for simplicity set $V_2=0$ in the following. A finite imaginary part of the nearest-neighbor bond hopping, $\chi_2={\rm Im }\langle c_i^{\dag}c_j\rangle\neq 0$, couples the two bands thereby opening a gap. In lowest order in $V_1$, the matrix describing this coupling is given by
\begin{equation}
H_{\rm QAH}({\boldsymbol k})=\begin{pmatrix}
\epsilon_2^*({\boldsymbol k})&2i\sqrt{3}V_1\chi_2\\
-2i\sqrt{3}V_1\chi_2&\epsilon_3^*({\boldsymbol k})
\end{pmatrix},
\label{eq:HQAH}
\end{equation}
where $\epsilon^*_{2,3}(\boldsymbol{k})$ is obtained from Eq.~\eqref{eq:kagome_disp} by replacing $t$ by
\begin{equation}
t^*=t+V_1\chi_0.
\label{eq:tstar}
\end{equation}
Here, $\chi_0=\langle c_i^{\dag}c_j\rangle_0=1/6$ denotes the nearest neighbor hopping expectation value in the noninteracting model and Eq.~\eqref{eq:tstar} takes into account the effect of the Fock term in lowest order in $V_1$. The self-consistency equation for $\chi_2$ reads
\begin{equation}
1=V_1\int_{2t^*-\Lambda/2}^{2t^*+\Lambda/2}\!\!d\epsilon\frac{N(\epsilon)}{\sqrt{(2t^*-\epsilon)^2+48V_1^2\chi_2^2}},
\label{eq:gap}
\end{equation}
where $\Lambda$ is a cutoff energy of the order of $t^*$ which is not accessible in the low energy description, and $N(\epsilon)$ is the noninteracting density of states. Solving Eq.~\eqref{eq:gap} for the order parameter $\chi_2$ yields
\begin{equation}
\chi_2=\frac{\Lambda}{2\sqrt{3}V_1}e^{-1/(V_1N_0)},
\label{eq:chi2}
\end{equation}
which holds for small values of the dimensionless coupling constant $V_1N_0$. Here, we have introduced the density of states at the QBCP:\cite{Indergand:2005}
\begin{equation}
N_0=N(2t^*)=\frac{\sqrt{3}}{2\pi t^*}.
\end{equation}
The gap is proportional to the order parameter $\chi_2$ and from the result Eq.~\eqref{eq:chi2} and the eigenvalues of Eq.~\eqref{eq:HQAH} it follows that
\begin{equation}
\Delta_{\rm QAH}=4\sqrt{3}V_1\chi_2=2\Lambda e^{-1/(V_1N_0)}.
\label{eq:gapQAH}
\end{equation}
We have checked that the exponential dependence given in Eq.~\eqref{eq:chi2} is indeed consistent with our full numerical evaluation of the self-consistency equations. A similar exponential dependence is also found in a one-loop renormalization group treatment~\cite{sun:2009} although the dimensionless coupling is renormalized compared to Eq.~\eqref{eq:gapQAH}.

\begin{figure}[tbp]
\centering
\includegraphics[width=0.7\linewidth]{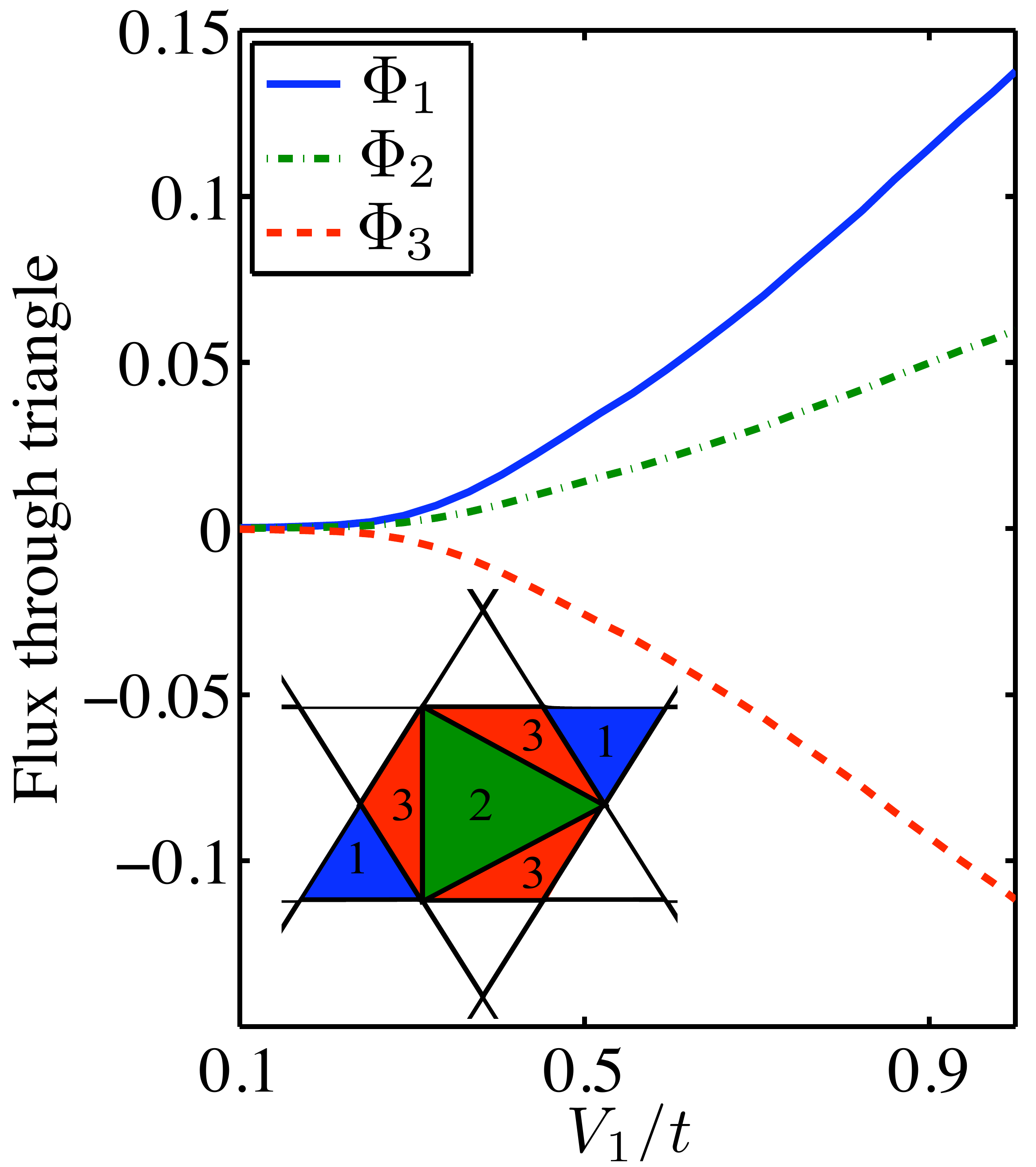}
\caption{(Color online) The fluxes $\Phi_{1,2,3}$ through elementary triangles in the QAH phase at filling fraction $f=2/3$. These elementary triangles form the unit cell, as shown in the inset. Because of the periodic boundary conditions on the unit cell, the net flux is zero and the individual fluxes satisfy $2\Phi_1+\Phi_2+3\Phi_3=0$. We have set $V_2=V_1/2$ and $V_3=0$.}
\label{fig:fluxes}
\end{figure}

We now come back to the general situation where both $V_1$ and $V_2$ are finite. In general, the QAH phase is driven by both a complex first and second neighbor hopping expectation value. Furthermore, one can define an explicit deformation\cite{Ruegg:prb10} of a tight-binding model with complex nearest neighbor hopping on the kagome lattice to show that its ground state is adiabatically connected to the ground state of a model with real nearest-neighbor hopping and only complex second nearest neighbor hopping. Therefore, the QAH phase generated by $V_2$ belongs to the same topological class as the one generated by $V_1$. Figure~\ref{fig:fluxes} shows the fluxes $\Phi_{1,2,3}$ through three elementary triangles forming the unit cell. In this figure, we set $V_2=V_1/2$. Because of the periodic boundary conditions on a unit cell, the fluxes satisfy $2\Phi_1+\Phi_2+3\Phi_3=0$.
Moreover, they are all finite indicating the presence of an imaginary hopping amplitude in both the first and the second neighbor effective hopping.

\subsubsection{CDW phases at $2/3$ filling}
\begin{figure}[tbp]
\centering
\includegraphics[width=0.9\linewidth]{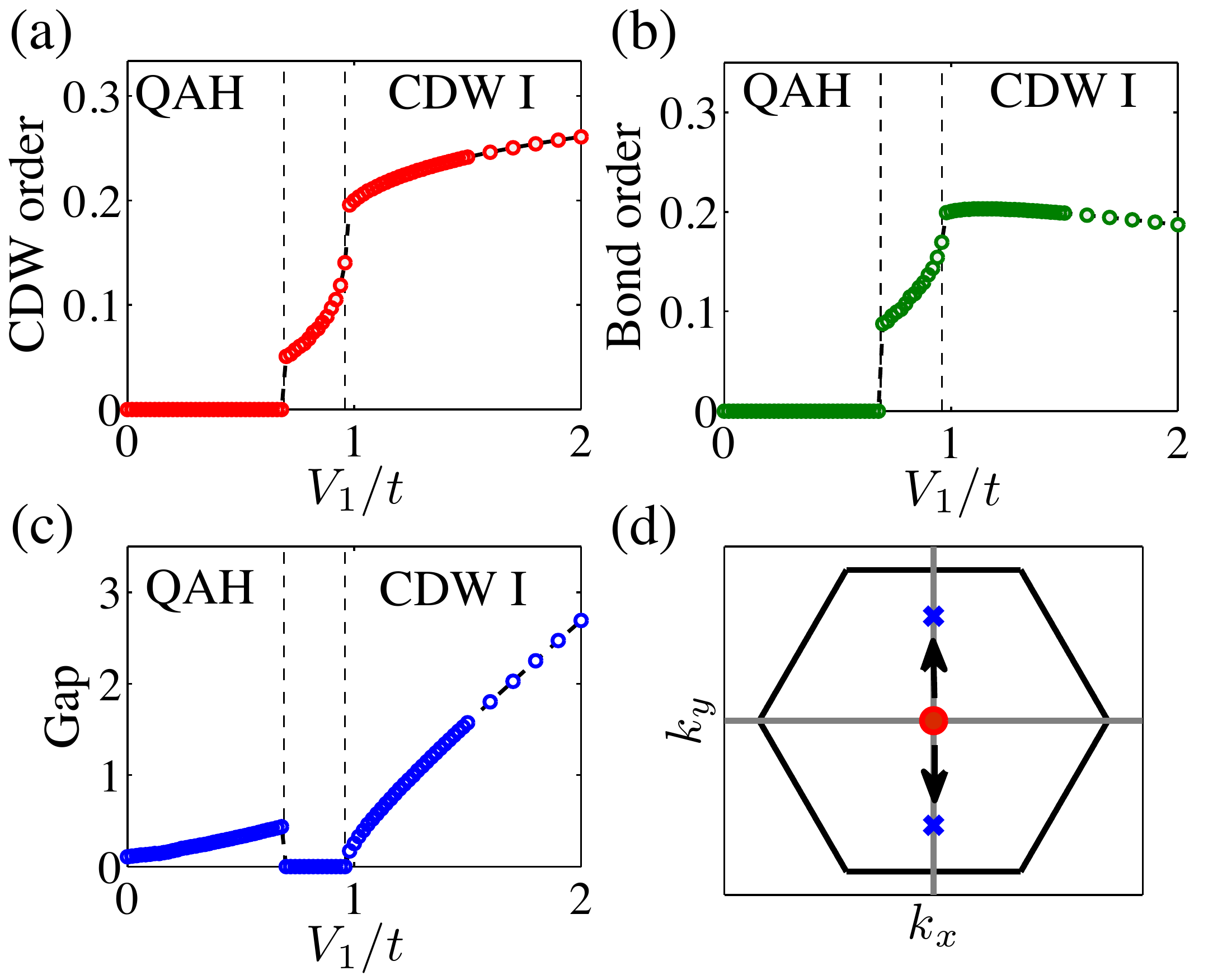}
\caption{(Color online)(a) the CDW order parameter $\rho$ (b) the bond order defined as the difference between two nearest strong and weak bonds (c) the gap at $2/3$ filling fraction on the kagome lattice. We have set $V_2=t$. (d) the splitting of a QBCP(red circle) into two Dirac points(blue cross) for gapless CDW~I. }
\label{fig:kagome_gap2}
\end{figure}

For intermediate to large interaction strengths, a CDW phase is stable. At $V_2=0$ and large $V_1$, the CDW~III phase has the lowest energy.
However, the difference in energy per site compared to CDW~I is only of the order $10^{-3}t$ and becomes smaller the bigger $V_1$. As a result, a very small but finite $V_2$ is sufficient to stabilize CDW~I over CDW~III. In contrast to the situation at filling fraction $f=1/3$, at $f=2/3$ CDW~III can not profit from a ``nesting" condition. The energy gain compared to CDW~I is therefore very small. At $V_2=0$, a first-order phase transition from a QAH state to a CDW~I (II) state takes place at $V_1\approx1.47t$. Numerically, we can not resolve any difference in the energy between CDW~I and II for $V_2=0$. Interestingly, there are nodes in the CDW~I (II) phase where the gap vanishes. The transition from the QAH phase to the CDW~I state with nodes is an example where a transition from a gapped phase (QAH) to a gapless phase (CDW~I) occurs by increasing the interaction strength, [see Fig.~\ref{fig:kagome_gap2}(c)].

The gapless CDW~I persists even for finite $V_2$. Figure~\ref{fig:kagome_gap2} shows the CDW order parameter $\rho$, the bond order $\nu=\chi_s-\chi_w$ defined as the difference between two nearest strong and weak bonds and the magnitude of the gap as a function of $V_1$ for fixed $V_2=t$. We can see that a finite CDW order is accompanied by a finite bond order. In fact, both types of orders jump to a finite value at the transition $V_{1c_1}$ from the QAH to the CDW~I phase. Note, however, that the gap is zero up to a second critical interaction strength $V_{1c_2}$ indicating the presence of band degeneracy points below $V_{1c_2}$. At $V_{1c_2}$ a kink is observed in the order parameters and the gap gradually starts to increase.

In the following, we show that the gapless CDW~I phase results from the splitting of the QBCP into two nodes.\cite{sun:2009} We notice that the change of bond-order is one order of magnitude smaller than $t$ and therefore can be neglected for the moment. We find the following low-energy Bloch Hamiltonian
\begin{equation}
H_{\rm \small CDW}(\boldsymbol{k})=
\begin{pmatrix}
\epsilon_2^*(k)\!+\!t^*u\cos(2\phi)&\frac{t^*u}{2}\sin(2\phi)\\
\frac{t^*u}{2}\sin(2\phi)&\epsilon_3^*(k)\!-\!t^*u\cos(2\phi)
\end{pmatrix},
\label{eq:HCDW}
\end{equation}
where $u=2(V_1+V_2)(\rho+3\nu)/t^*$. The renormalized hopping $t^*$ is given by Eq.~\eqref{eq:tstar} with $\chi_0$ replaced by $\bar{\chi}=(\chi_s+2\chi_w)/3$ and we have neglected the Fock-terms generated by the $V_2$ interaction (this term is negligible in practice). In Eq.~\eqref{eq:HCDW} we have introduced polar coordinates $(k,\phi)$ which are defined by ${\boldsymbol k}\cdot \boldsymbol{e}_n=k\cos\phi$. Note that right at the $\Gamma$-point the angle $\phi$ is not well-defined. Equation~\eqref{eq:HCDW} should be contrasted with Eq.~\eqref{eq:HQAH} for the QAH phase: as opposed to the QAH order parameter, the CDW order parameter introduces an anisotropic {\it angle dependent} effective coupling between the two bands. Expanding the dispersion around the $\boldsymbol{\Gamma}$ point we find for the eigenvalues of Eq.~\eqref{eq:HCDW},
\begin{eqnarray}
E_2(k,\phi)/t^*&=&2-[k^2+\sqrt{B(k,\phi)}]/8,\\
E_3(k,\phi)/t^*&=&2-[k^2-\sqrt{B(k,\phi)}]/8.
\end{eqnarray}
The function $B(k,\phi)$ is given by
\begin{equation}
B(k,\phi)=k^4-16k^2u\cos(2\phi)+64u^2.
\end{equation}
It has roots at two points where the two bands touches:
\begin{equation}
k_u=\sqrt{8u},\quad\phi=0,\pi.
\end{equation}
This analysis shows that a finite CDW~I order splits the QBCP into two nodes moving along the line defined by the vector $\boldsymbol{e}_n$. The bottom right panel of Fig.~\ref{fig:kagome_gap2} illustrates the situation for $\boldsymbol{e}_3$. We have calculated the Berry phase (winding number)\cite{sun:2009} of the QBCP and found that it is $2\pi$ ($=0$ mod $2\pi$).  The corresponding Berry phases (winding numbers) of the two nodes appearing in the gapless CDW~I are both $\pi$. Thus, the QBCP does splits into two Dirac points with Berry phases $\pi$ conserving to total winding number, as it was suggested in Ref.~[\onlinecite{sun:2009}].

\subsection{Comparison with existing work}

Recently, several numerical works\cite{nishimoto:2010,OBrien:2010} appeared dealing with the charge density wave order on the kagome lattice at $f=1/3$ or $f=2/3$. Here, we want to briefly relate our results with their findings. In Nishimoto $et.al.$'s work,\cite{nishimoto:2010} the authors considered the large $V_1$ limit with vanishing $V_2$. They showed that CDW~III is the ground state that is consistent with the ``plaquette" state obtained from an effective quantum dimer model on the honeycomb lattice.~\cite{Moessner:2001} In this strong interacting limit, $f=1/3$ and $f=2/3$ are equivalent and numerical calculations\cite{nishimoto:2010} confirm that CDW~III is stabilized by the ring exchange process proportional to $|t|^3/V^2$. Interestingly, fractionalized excitations with charge $e/2$ have recently also been reported in the strong coupling limit.\cite{OBrien:2010}

Our mean-field calculation cannot capture the resonating nature of the quantum dimer model and is not valid in the strongly interacting limit. However, CDW~III in the mean-field treatment can be viewed as the ``classical" configuration of plaquette states. At $1/3$ filling fraction, the CDW~III is found to be more stable than either CDW II or I at large $V_1$, and the energy difference between them becomes smaller as $V_1$ grows. This is consistent with Nishimoto $et.al.$'s work. Furthermore, we predict a metal-insulator transition takes places at $V_{1c}=3.1t$ at $1/3$ filling, which is in quite good agreement with their result $V_{1c}=4.0t$. However, at $2/3$ filling, our mean-field results differ significantly from theirs. We find that the leading instability at small interactions is the QAH state that spontaneously breaks time-reversal symmetry and has an exponentially small gap.~\cite{sun:2009} A metal-insulator transition takes place around $V_{1c}=2t$ in our study while Nishimoto $et. al.$ reported a metal-insulator transition at finite $V_{1c}=2.6t$.

\section{Spinful model on the kagome lattice}
Let us now turn to the spinful model on the kagome lattice.
The additional spin degrees of freedom add considerable complexity to
the problem and introduce several more potential phases.

\subsection{Candidate phases}
The SM and the CDW phase are equivalent to those in the spinless model. Here, we discuss additional phases which appear in the spinful model.
\subsubsection{Topological insulator and quantum anomalous Hall state}
The topological insulator and quantum anomalous Hall state are both stabilized by a complex Fock term of nearest or second neighbor interaction which gives rise to a complex hopping amplitude. The difference between TI and QAH phases can be described by the $2\times 2$ matrix $\langle c_{i\alpha}^{\dag}c_{j\beta}\rangle$ defined in the spin space as discussed below.

The QAH state breaks the time reversal symmetry but not the spin rotation SU(2) symmetry. Therefore, the most general form of the uniform phase consistent with these requirements is
\begin{equation}
\langle c_{i\alpha}^{\dag}c_{j\beta}^{}\rangle=\left[(\chi_{1}+i\chi_2)\sigma_0\right]_{\alpha\beta},
\end{equation}
where $\chi_1$ and $\chi_2$ are real numbers and $\sigma_0=\hat{1}$ is the identity matrix. A phase with a finite $\chi_2$ shows an anomalous quantum Hall effect and a non-zero Chern number.

On the other hand, the TI does not break time reversal symmetry but breaks the SU(2) spin rotation symmetry down to U(1). The most general form is therefore
\begin{equation}
\langle c_{i\alpha}^{\dag}c_{j\beta}^{}\rangle=\left[\chi_{1}\sigma_0+i\chi_2(\vec{n}\cdot\vec
{\sigma})\right]_{\alpha\beta},
\end{equation}
where $\chi_1$ and $\chi_2$ are both real numbers and $\vec{n}$ is a unit vector describing how the SU(2) spin rotation symmetry is broken. In other words, spin-rotation symmetry is only preserved for rotations around $\vec{n}$. Without loss of generality we can assume $\vec{n}=\hat{n}_z$. We note that allowing $\chi_1$, $\chi_2$ and $\vec{n}$ to be spatially dependent allows one to study topological defects of the order parameter, such as skyrmions, providing a potential route for exotic superconductivity.\cite{Grover:2008} On the other hand, in contrast to their two-dimensional counterpart, three-dimensional interaction-driven TIs completely break the spin-rotation symmetry and their order parameter involves a rotation matrix. Again, it is possible to study topological defects which host protected modes.\cite{Yi:2009}

A short inspection of the mean-field free energy of the TI and the QAH phase shows that these two phases are degenerate on the mean-field level. It is likely that fluctuations around the mean-field state might favor one phase over the other. Because the TI breaks the continuous spin-rotation symmetry, there are Goldstone modes in the ordered phase.\cite{Raghu:2008} It was suggested\cite{Raghu:2008} that quantum fluctuations associated with these modes lower the ground state energy of the TI as compared with the QAH phase which does not have Goldstone modes.  This argument appears to be confirmed via ``unbiased" functional renormalization group methods.\cite{Raghu:2008}  We don't see any reason for those arguments not to hold in the present case as well.
\subsubsection{Spin-charge-density waves}
There is another class of phases that emerges as a result of the special filling fractions, the non-bipartite nature of the kagome lattice and the additional spin degrees of freedom. We term it ``spin charge density wave" (SCDW) because it involves both a spin and a charge density wave. In our mean-field calculations we restrict to phases which do not break the translational symmetry. By solving self-consistency equations we identify two types of SCDWs which are stable for some interaction parameters.

\begin{figure}[tbp]
\centering
\includegraphics[width=0.75\linewidth]{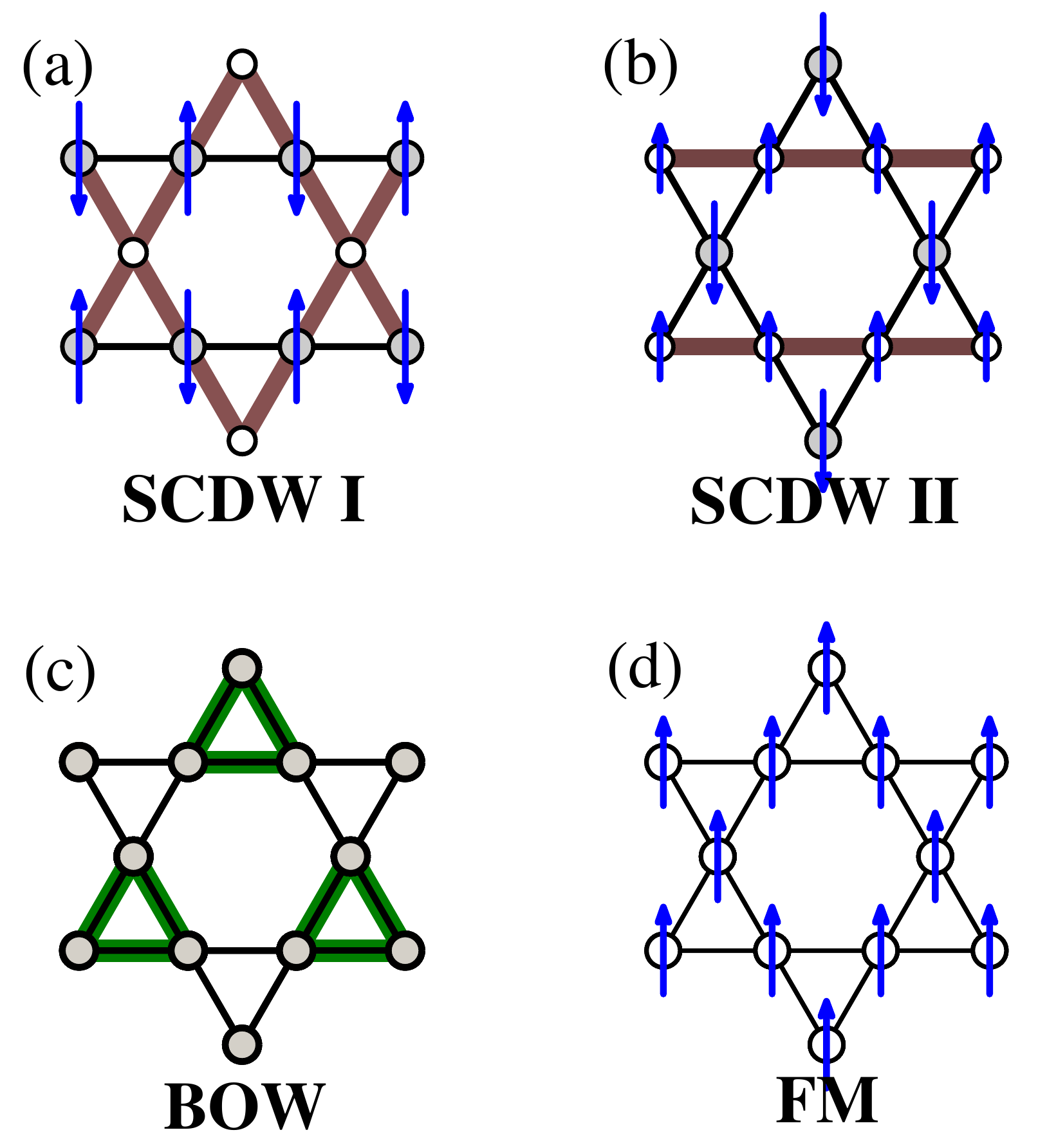}
\caption{(Color online) Schematic of four types of candidate phases on the kagome lattice for $1/3$ filling fraction: (a) SCDW~I (b) SCDW~II (c) bond-order wave (BOW) and (d) ferromagnet (FM). Upward arrows and downward arrows denote the magnetization on each site. The same/different circles represent same/different numbers of fermions on corresponding sites. For simplicity we only show nearest bonds (the addition of two spin-resolved bonds) and do not show the second nearest bonds.  Stronger bonds are shown in bold.}
\label{fig:kagome_SCDW_pattern}
\end{figure}

The first pattern, SCDW~I, is characterized by the following distribution:
\begin{subequations}
\label{eqn:SCDWI}
\begin{align}
\langle n_{1\uparrow}\rangle&=f+\rho+m,\\
\langle n_{1\downarrow}\rangle&=f+\rho-m,\\
\langle n_{2\uparrow}\rangle&=f+\rho-m,\\
\langle n_{2\downarrow}\rangle&=f+\rho+m,\\
\langle n_{3\uparrow}\rangle&=f-2\rho,\\
\langle n_{3\downarrow}\rangle&=f-2\rho.
\end{align}
\end{subequations}
Here, $\rho$ and $m$ are the charge density and spin density order parameter, respectively. Furthermore, we assume that the symmetry of the bond expectation values is determined by the symmetry of the spin-charge configuration and therefore, three different spin-resolved bond expectation values have to be introduced. The phase SCDW~I is schematically shown in Fig.~\ref{fig:kagome_SCDW_pattern}(a).

The other configuration, SCDW~II, is characterized by the following distribution:
\begin{subequations}
\label{eqn:SCDWII}
\begin{align}
\langle n_{1\uparrow}\rangle&=f-\rho+m,\\
\langle n_{1\downarrow}\rangle&=f-\rho-m,\\
\langle n_{2\uparrow}\rangle&=f-\rho+m,\\
\langle n_{2\downarrow}\rangle&=f-\rho-m,\\
\langle n_{3\uparrow}\rangle&=f+2\rho-2m,\\
\langle n_{3\downarrow}\rangle&=f+2\rho+2m.
\end{align}
\end{subequations}
The schematics of the SCDW~II is shown in Fig.~\ref{fig:kagome_SCDW_pattern}(b). In addition, we also introduce four different spin-resolved bond expectation values to make it consistent with the above spin-charge distribution.

Both SCDWs have zero magnetization in the unit cell. However, they differ in that SCDW~I has antiferromagnetic order in the $\boldsymbol{a_1}$ direction,  ferromagnetic order in the $\boldsymbol{a_2}$ and $\boldsymbol{a_1}-\boldsymbol{a_2}$ directions, while SCDW~II has ferromagnetic ordering in the $\boldsymbol{a_1}$ direction but antiferromagnetic ordering in the $\boldsymbol{a_2}$ and $\boldsymbol{a_1}-\boldsymbol{a_2}$ directions.

In the next section, we will see that SCDWs arise in the case of large $U$ but small or moderate $V_1$ and $V_2$. At $1/3$ filling fraction, SCDWs can be understood as a means to reduce the on-site interactions by single occupancy at two sites in a unit cell. Therefore they become unstable when $V_1$ or $V_2$ becomes large and CDW dominates. We stress that the solutions of SCDWs are saddle points of the free energy instead of a global minimum in the usual situation, therefore, one has to solve self-consistency equations directly to obtain the SCDW solutions.

\subsubsection{Bond-order wave}
Next we consider the bond-order wave (BOW) as has been found in Ref.~[\onlinecite{Indergand:2006}] for the $t$-$J$ model at $f=1/3$ under quite general conditions. The BOW is characterized by a uniform charge distribution and a bond order which breaks the inversion symmetry of the unit cell by establishing strong bonds $\chi_s$ for the up triangles and weak bonds for the down triangles $\chi_w$.  It is schematically shown in Fig.~\ref{fig:kagome_SCDW_pattern}(c).

\subsubsection{Ferromagnet}
The ferromagnetic state (FM) is characterized by a uniform magnetization density $m$. The spin densities are given by
\begin{equation}
\langle n_{i\uparrow}\rangle=f+m/2,\quad \langle n_{i\downarrow}\rangle=f-m/2,
\label{eq:nsFM}
\end{equation}
and we introduce the Fock terms
\begin{equation}
\chi_{\uparrow}=\langle c^{\dag}_{i\uparrow}c_{j\uparrow}^{}\rangle, \quad \chi_{\downarrow}=
\langle  c^{\dag}_{i\downarrow}c_{j\downarrow}^{}\rangle.
\label{eq:chiFM}
\end{equation}
The Fock terms are different for nearest-neighbor and second-neighbor bonds. Equations~\eqref{eq:nsFM} and \eqref{eq:chiFM} are used as an {\it ansatz} to solve the self-consistency equations numerically. A finite magnetization of the form Eq.~\eqref{eq:nsFM} introduces a Zeeman field which uniformly lowers the energy of the spin-$\uparrow$ electrons with respect to the spin-$\downarrow$ electrons by $Um$. At $f=1/3$ the maximally polarized state is obtained when there are two up electrons per unit cell. At $f=2/3$, the maximally polarized state corresponds to 3 up electrons and one down electron per unit cell. At both filling fractions the saturated value of the magnetization is $m_{\rm sat}=2/3$. In the next section, we will see that the maximally polarized FM state arises in the large $U$ limit.

\subsection{Phase diagrams at $1/3$ filling}
We first discuss the role of $U$ and $V_1$ and set $V_2=V_3=0$. This allows for a direct comparison with the phase diagram at 2/3 filling shown below.
\subsubsection{$U$-$V_1$ phase diagram at 1/3 filling}
\begin{figure}[tbp]
\centering
\includegraphics[width=0.9\linewidth]{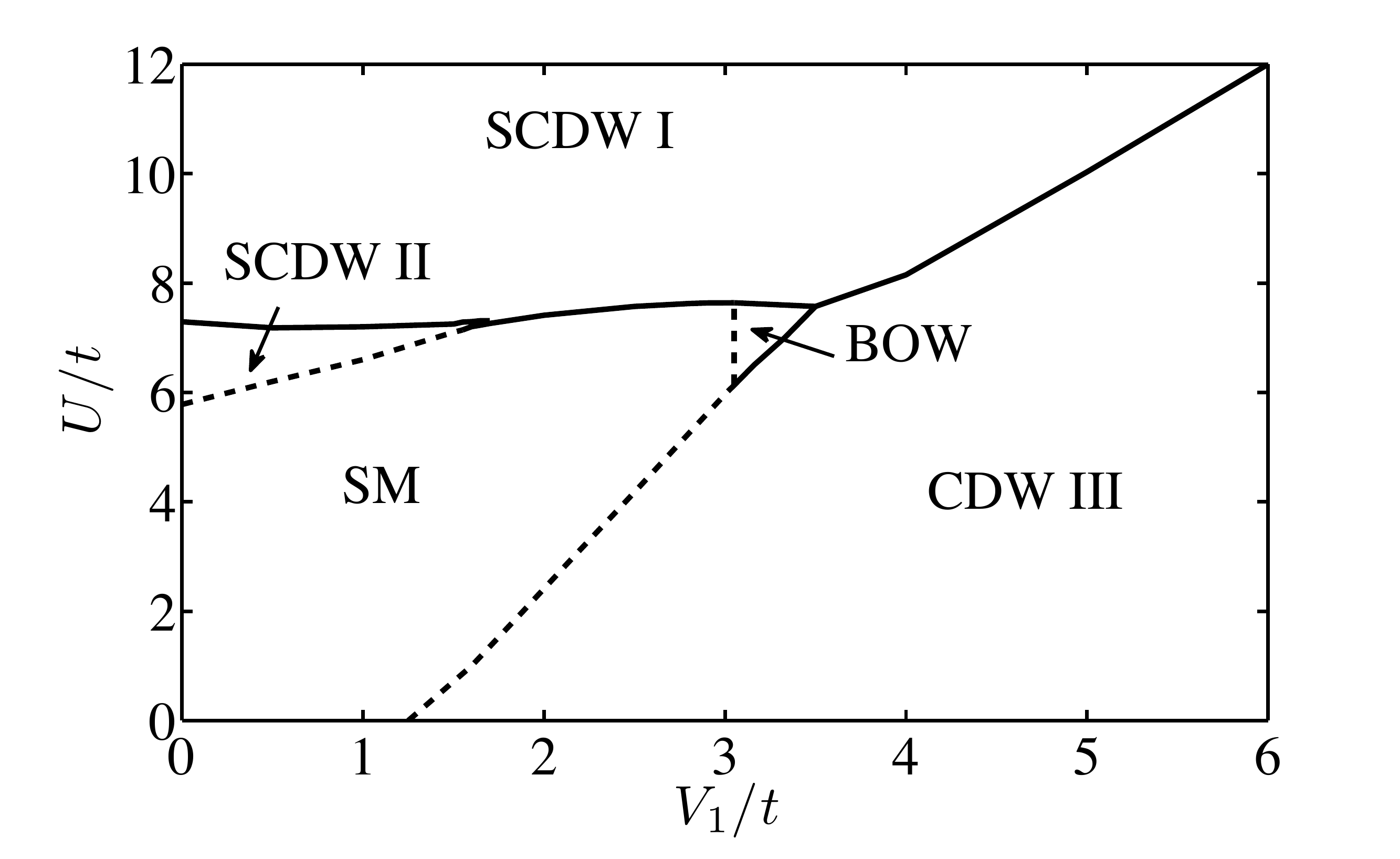}
\caption{The phase diagram of the spinful model at 1/3 filling on the kagome lattice. The SCDW I and II phase involve both a finite charge and spin density wave order parameter. Furthermore, when $U$ competes with $V_1$ a bond-order wave (BOW) is found. Solid lines indicate first order and dashed lines second order transitions.}
\label{fig:phasediagram1o3}
\end{figure}

The $U$-$V_1$-phase diagram is shown in Fig.~\ref{fig:phasediagram1o3}. Similar to the spinless model, we find that the SM is stable for small to intermediate interactions which we again attribute to the vanishing density of states at the Fermi energy in the noninteracting limit. For dominant $V_1$ interaction, we find that the CDW~III is stable and the transition from SM to the CDW~III is second order. For dominant onsite interaction $U$, a SCDW phase is stabilized. Both patterns SCDW~I and SCDW~II are stable for some values of the interaction. We note that for small $V_1$ there is a second order transition from the SM to the SCDW~II with nodes.

For intermediate $U$ and $V_1$ we find BOW is the favored ground state on kagome lattice at 1/3 filling.\cite{Indergand:2006} It requires that $U$ is of the same order as $V_1$ to suppress the CDW~III. On the other hand, it requires a reasonable value of $V_1$ to generate the bond order at all. However, we expect that the superexchange mechanism (second order in $t/U$), which is not captured in our mean-field treatment, could stabilize this phase also for smaller $V_1$.\cite{Indergand:2006}

At quite large onsite interactions ($U\sim 20t$) a FM phase is stabilized (not shown). The FM state is fully polarized at the mean-field level and has an energy gain of
\begin{equation}
e_{\rm FM}-e_{\rm SM}=\bar{\epsilon}_2-\bar{\epsilon}_1+\frac{V}{24t}(\bar{\epsilon}_1^2-\bar{\epsilon}_2^2-2\bar{\epsilon}_1\bar{\epsilon}_2)-\frac{U}{3}
\label{eq:eFM1o3}
\end{equation}
per unit cell as compared to the SM phase. In Eq.~\eqref{eq:eFM1o3}, we have introduced the average kinetic energy of the filled band $n$,
\begin{equation}
\bar{\epsilon}_n=\frac{1}{N}\sum_{\boldsymbol{k}}\epsilon_n(\boldsymbol{k}),
\label{eq:ekin}
\end{equation}
where $N$ is the number of unit cells in the lattice,  and the dispersion relation $\epsilon_{n}(\boldsymbol{k})$ is given in Eq.~\eqref{eq:kagome_disp}.
We note that the presence of a FM state for large interactions is consistent with numerical studies.\cite{Pollmann:2008}
Finally, we note that the QAH/TI phase does not occur in the absent of a finite $V_2$. Again, this is in agreement with the spinless case.

\subsubsection{$U$-$V_2$ phase diagram at $1/3$ filling}
\begin{figure}[tbp]
\centering
\includegraphics[width=0.9\linewidth]{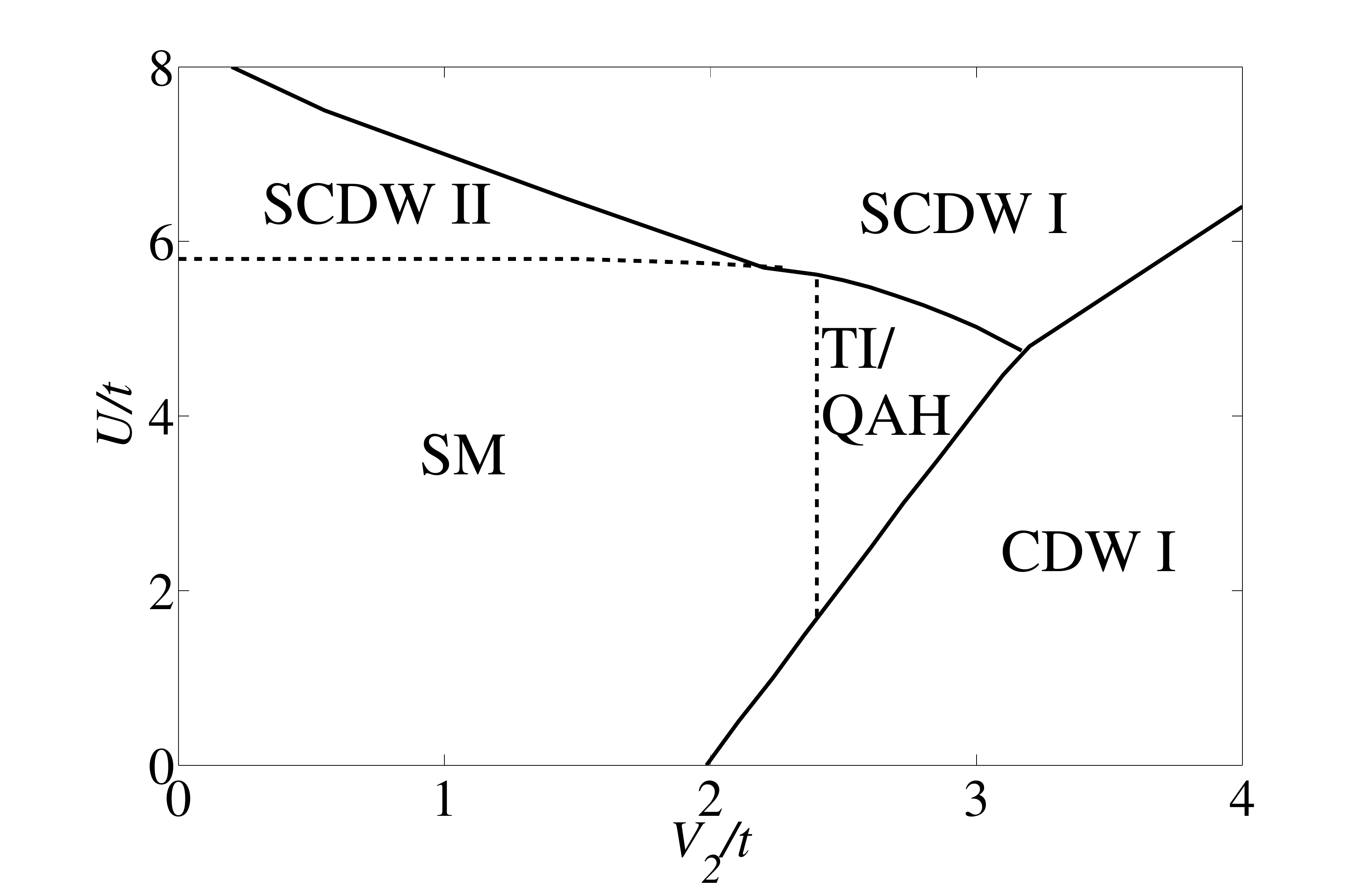}
\caption{The $U$-$V_2$ phase diagram for $V_1=0$ and $V_3=0.4t$. An interaction-driven TI appears for finite $U$ and $V_2$. Solid lines indicate first order and dashed lines second order transitions}
\label{fig:phasediagram1o3V2}
\end{figure}

The $U$-$V_2$ phase diagram of is shown in Fig.~\ref{fig:phasediagram1o3V2}. Like in the spinless case at $1/3$ filling fraction, we add a small $V_3$ interaction to suppress CDW I for finite $V_2$ and stabilize TI/QAH. The overall structure is quite similar to the $U$-$V_1$ phase diagram. However, the charge density wave has pattern I for large $V_2$ since large $V_2$ does not favor CDW III but CDW I. BOW phase is now replaced by the QAH/TI phase. That the topological phase appears in the middle of the phase space
seems to be a rather universal feature in systems which have a Dirac
point and has also been reported on the honeycomb and the diamond lattice.\cite{Raghu:2008,Yi:2009} For large $U\sim6t$, SCDW II is stabilized and we find that it is gapless. A first-order phase transition from SCDW II to SCDW I occurs when $V_2$ increases and finally CDW I dominates for large $V_1$. Note a FM state occurs at even larger $U\sim 20t$ (not shown).

\subsection{Phase diagram at $2/3$ filling}
\begin{figure}[tbp]
\centering
\includegraphics[width=0.9\linewidth]{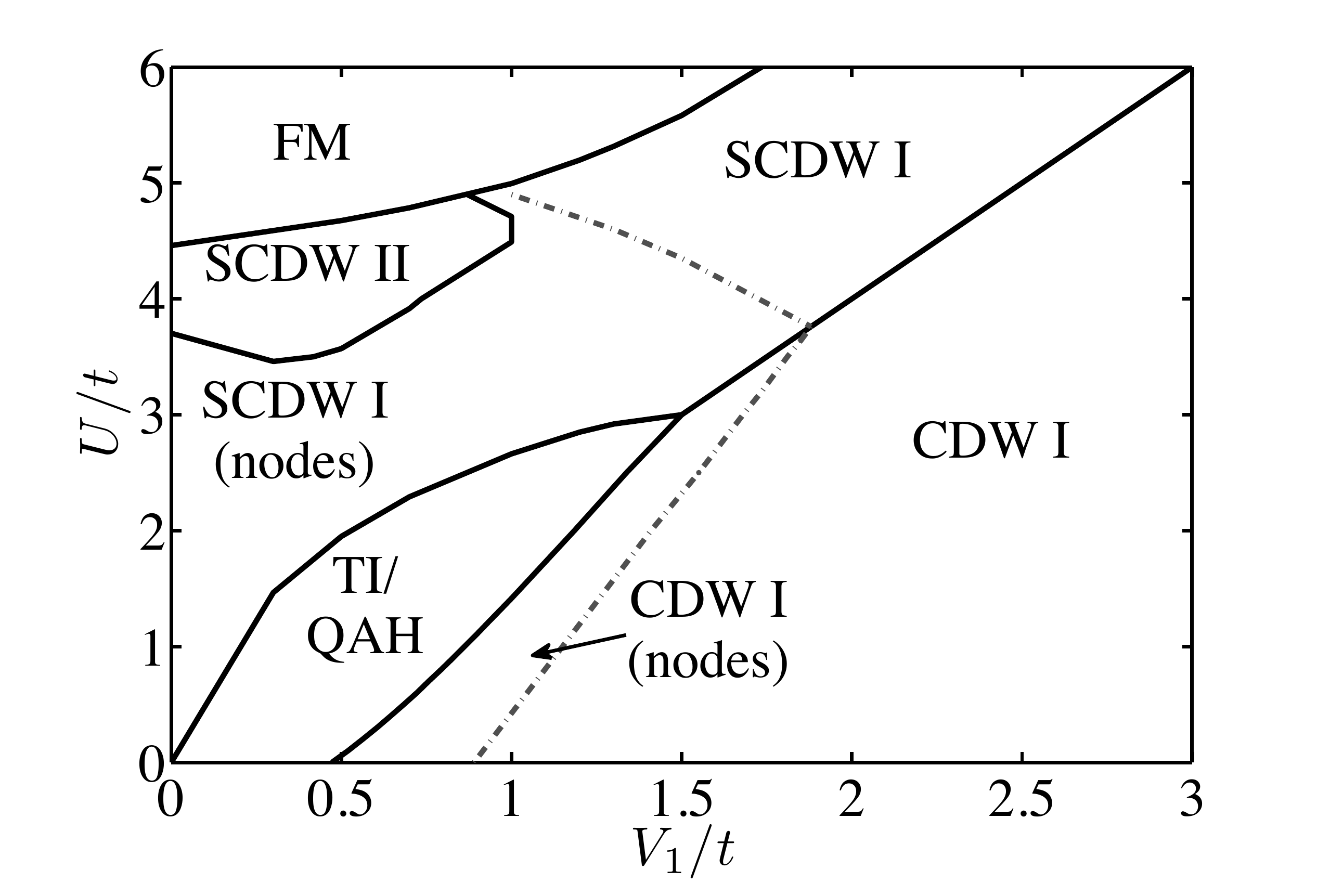}
\caption{The $U-V_1$ phase diagram for $V_2=V_3=0$ at $2/3$ filling fraction. Similar to the spinless case at $2/3$ filling fraction, CDW I has nodes which separates itself from gapped phase by a dash-dot line. Similarly, SCDW I has nodes for small interaction strengths.}
\label{fig:phasediagram2o3}
\end{figure}
For 2/3 filling we focus only on the $U$-$V_1$ phase-diagram. The phase diagram is shown in Fig.~\ref{fig:phasediagram2o3}. Most importantly, we found that the dominant instability for arbitrarily small $V_1$ is to the QAH/TI phase and this phase survives also for finite $U$. Increasing $V_1$ further, there is a first order transition to a gapless and then gapped CDW.

We note that for $V_1\lesssim 0.3t$ the energy difference between various states is very small: SCDW~I and II as well as QAH/TI have energy differences of less than $10^{-6}t$ per unit cell and we had to use a very high precision in the numerical calculation to resolve the phase diagram. However, for larger values of $U$ the FM phase is clearly favored in the mean-field calculation.
This is again a maximally polarized FM state. The energy density as compared to the SM phase is
\begin{equation}
e_{\rm FM}-e_0=-(\bar{\epsilon}_1+2\bar{\epsilon}_2)-\frac{V_1}{24t}(\bar{\epsilon}_1^2-\bar{\epsilon}_2^2-2\bar{\epsilon}_1\bar{\epsilon}_2)-\frac{U}{3},
\label{eq:eFM2o3}
\end{equation}
and we have used the definition Eq.~\eqref{eq:ekin}. While the energy gain for the onsite repulsion $U$ is the same in the SCDWs, it is the kinetic energy which favors the FM phase over the SCDWs for large $U$.

\section{Spinless Fermions on Decorated Honeycomb Lattice}

In this section, we briefly examine the possibility of an interaction-driven QAH state for spinless fermions on the decorated honeycomb lattice and discuss the relationship of the QAH phase with other competing phases. The study of interaction effects on this lattice is partially motivated by a recent paper that exactly solved\cite{Yao:2007} the Kitaev model on this lattice in the strongly interacting limit of the underlying fermions;  our previous paper established the existence of TIs on this lattice in the noninteracting limit.\cite{Ruegg:prb10} One natural question to ask is what will happen for intermediate interaction strengths where the Hartree-Fock mean-field approximation is still valid. We work at half filling and $t'=t$ and show that the QAH state is the leading instability in the presence of interactions.  Moreover, it occupies a rather wide region in the phase diagram.

\begin{figure}[tbp]
\centering
\includegraphics[width=0.9\linewidth]{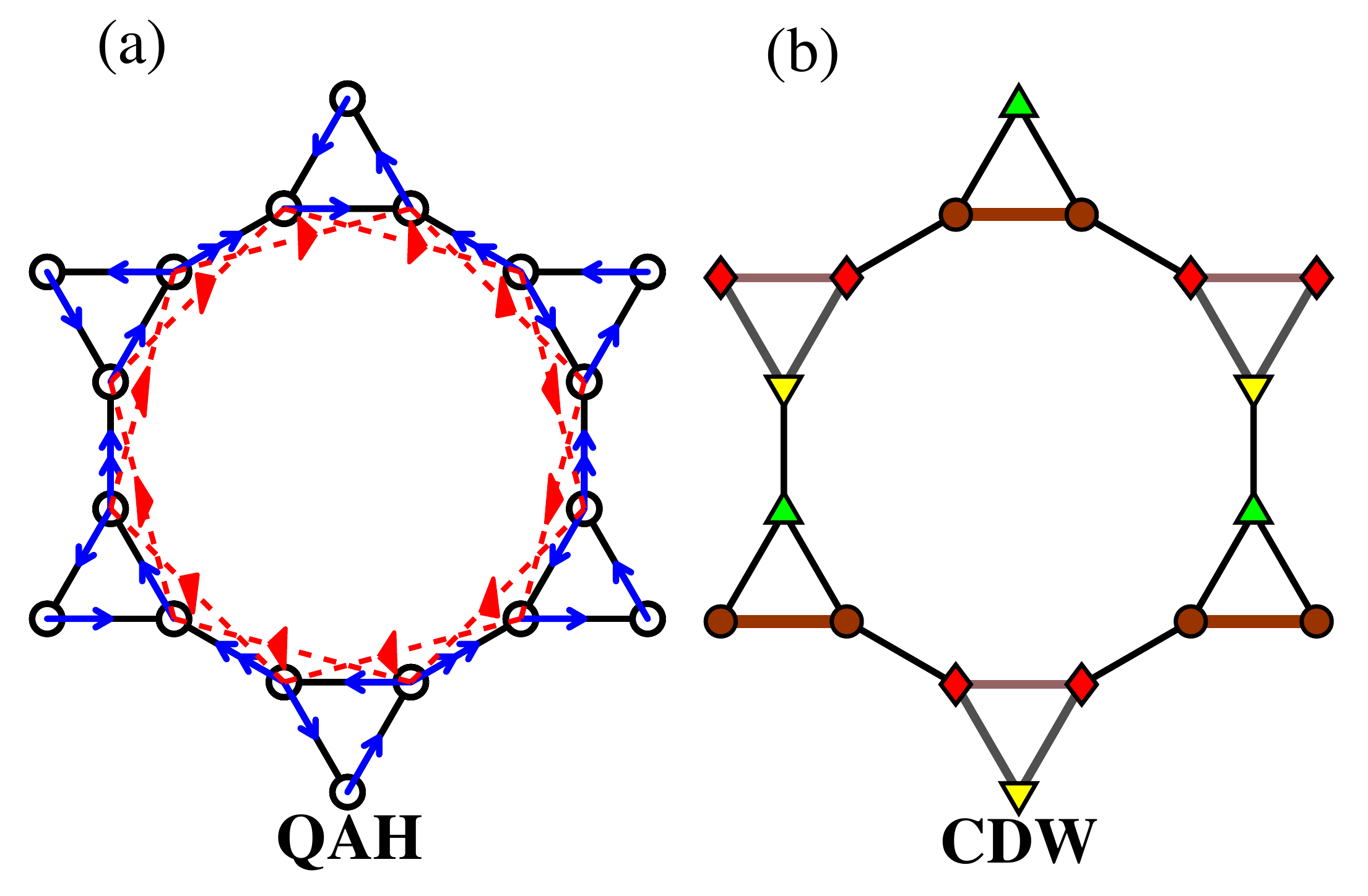}
\caption{(Color online) (a) the flux pattern developed by interactions that preserves the lattice symmetry but spontaneously breaks time reversal symmetry on decorated honeycomb (``star") lattice. The blue solid line with an arrow (two arrows) represents a nearest neighbor intra-triangle(inter-triangle) complex hopping while the red dash line with an arrow represents the second neighbor complex hopping. (b) the favorable CDW pattern from solutions of self-consistency equations. We have used different combinations of color and markers to show the mirror symmetry. Real first neighbor bonds consistent with symmetry of the CDW pattern have been assumed (second neighbor bonds not shown). }
\label{fig:DH_QAH_CDW_structure}
\end{figure}

The Fermi surface at $1/2$ filling lies at a quadratic band crossing point in the center of the Brillouin zone, where a flat band crosses a quadratic band. It allows the emergence of a QAH phase quite easily without any fine tuning of interaction strengths. We consider the nearest-neighbor interaction $V_1$ and second nearest-neighbor interaction $V_2$ on this lattice. For the $V_1$ interaction, we can introduce a dynamically generated flux pattern in the two triangles.  We also introduce a second-neighbor flux in the same way as in the kagome lattice [see Fig.~\ref{fig:DH_QAH_CDW_structure}(a)]. One key difference, however, is that we have to allow the possibility of different values of inter-triangle complex hopping parameters and intra-triangle complex hopping parameters due to nonequivalence of the two hopping parameters in the non-interacting limit.  One can easily show that if the flux through a unit cell is zero and time-reversal symmetry is broken, a QAH state is realized similar to the one on kagome lattice. In our calculation, we also find it is possible to have a BOW state if the phase of the flux is zero or $\pi$. The BOW is very close in energy to the QAH state ($10^{-6}t$), but appears to lose out for the parameter ranges we studied.

We will restrict ourselves to $\boldsymbol{q}=0$ CDW that originates from the Hartree term of the mean-field Hamiltonian. At $1/2$ filling, one can see that $V_1$ and $V_2$ frustrate each other, in contrast to the kagome lattice at $1/3$ filling where $V_1$ and $V_2$ both stabilize a CDW state. The CDW pattern we found based on the mean-field self-consistency equations is shown in Fig~\ref{fig:DH_QAH_CDW_structure}(b). Among several CDW solutions we have found, we identify this particular CDW with a mirror symmetry as having the lowest ground state energy. Real bond orders from the Fock term have also been introduced implicitly to be consistent with this CDW pattern.

\begin{figure}[tbp]
\centering
\includegraphics[width=0.8\linewidth]{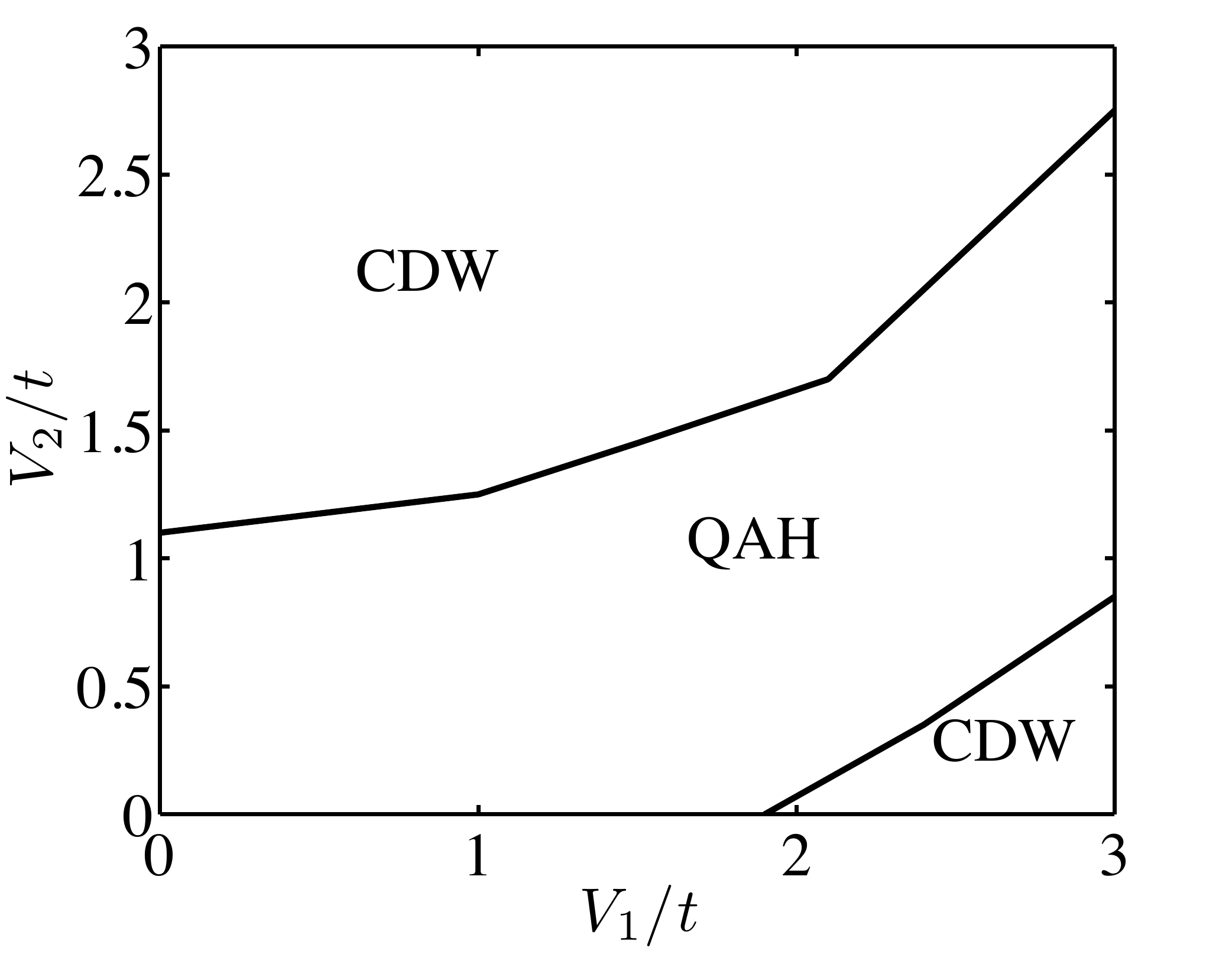}
\caption{(Color online) The phase diagram of the extended Hubbard model of spinless fermions at 1/2 filling on the decorated honeycomb lattice.  Due to the mutual frustration of $V_1$ and $V_2$, the QAH phase occupies the middle part of phase diagram and the regions of CDW phase have been split into two parts. We have set $t'=t$.}
\label{fig:DH_phase}
\end{figure}

The phase diagram is shown in Fig.~\ref{fig:DH_phase}. Along the horizontal axis where $V_2=0$, a BOW phase competes with the QAH for small interaction but is higher in energy by $\sim 10^{-6}t$.  A CDW phase develops for $V_1>1.9t$.
Generally, the energy gain from forming a uniform distribution of fermions and complex bonds between neighbor sites is very small compared to that in the CDW phases. Therefore, the QAH phase is the only  favorable ground state when either (i) there are no CDW solutions, or (ii) the CDW states are frustrated or suppressed and therefore have much higher ground state energy. The first case has been seen in the kagome lattice at $2/3$ filling fraction for small interactions, and the second situation is realized in the present case of the decorated honeycomb lattice at 1/2 filling where a QAH phase occupies a large region of the phase diagram. This is intimately related to the mutual frustration between the $V_1$ and $V_2$ interaction at $1/2$ filling. To see this, let us consider the large $V_2\gg V_1$ limit. One finds that in order to minimize the interaction $V_2$, the preferred ground state is the CDW configuration where three sites in a top triangle are almost occupied while the three sites in the bottom triangle in the same unit cell are almost empty. This is exactly the sublattice potential perturbation considered in Ref.~[\onlinecite{Ruegg:prb10}] that destabilizes the quantum spin Hall phase. Though this sublattice potential appears artificial at first sight, we show here that it can result from a many-body interaction. Clearly this configuration is not stable if $V_1$ is increased beyond a critical value. This explains the fact that at large $V_1$ or $V_2$ the CDW phase is the ground state, while the QAH state is the ground state when $V_1$ is comparable to $V_2$.

It is possible to perform a similar mean-field calculation for the spinful case, and one expects that a TI/QAH phase will dominate at small interactions strengths for 1/2 filling. However, the details are beyond the scope of this paper and is left to future work.

\section{Conclusions and Summary}
We have presented comprehensive Hartree-Fock mean-field calculations of the phase diagram for spinless and spinful fermions described by the extended Hubbard model on the kagome lattice and decorated honeycomb lattice. We have established the existence of interaction-driven topological phases at filling fractions where either Dirac points or quadratic band crossing points are involved. We find that both TI and QAH phases can be described by conventional complex bond order parameters. Quite generally, we find that at $2/3$ filling on the kagome lattice and 1/2 filing on the decorated honeycomb lattice (where a quadratic band crossing point is involved in the non-interacting limit), the TI/QAH phase is the leading instability for small interaction strengths. We have observed also that interaction-driven topological phases only exist beyond a critical interaction value when the Fermi surface lies at Dirac points at $1/3$ filling on kagome lattice (in the zero interaction limit). Furthermore, we discuss in detail various other phases including charge density wave, spin-charge density wave, bond-ordered wave, and ferromagnets on the two lattices.

An important lesson drawn from this study is that systems whose non-interacting band structures involve quadratic band crossing points can be unstable to topological phases with arbitrarily weak interactions, even in the absence of microscopic spin-orbit coupling.  We hope this work will aid in the search for topological states of matter by enlarging the class of candidate materials to include those which {\em do not} have strong microscopic spin-orbit coupling but do have certain features (such as quadratic band touching points) in their non-interacting band structure.

\begin{acknowledgments}
We appreciate stimulating discussions with Marcel Franz, Mike Hermele, Srinivas Raghu, Kai Sun, and Fan Zhang. We gratefully acknowledge financial support from ARO Grant W911NF-09-1-0527.
\end{acknowledgments}

\bibliography{MeanFieldPaper-5-20}

\end{document}